\documentstyle[A4,12pt]{article}
\newcounter{sct}

\begin{document}
\large
\pagestyle{headings}
\begin{center}
 {\bf
  {\Large IRREDUCIBLE GAUGE THEORIES IN THE\\ FRAMEWORK OF THE
Sp(2)-COVARIANT\\
          [1mm]
          QUANTIZATION METHOD}}\\[2mm]
              P. M. LAVROV\footnote{E-mail lavrov@tspi.tomsk.su}\\
 {\footnotesize{\it Tomsk State Pedagogical Institute, Tomsk, 634041,
Russia}}\\
           P. YU. MOSHIN {\small and} A. A. RESHETNYAK\\
 {\footnotesize{\it Tomsk State University, Tomsk, 634050, Russia}}\\
\end{center}
\begin{quotation}
{\footnotesize
        Irreducible gauge theories in both the Lagrangian and Hamiltonian
        versions of the Sp(2)-covariant quantization method are studied.
        Solutions to generating equations are obtained in the form of
        expansions in power series of ghost and auxiliary variables up to the
        3d order inclusively.}
\end{quotation}
{\bf 1. Introduction}\\[0.1mm]

 The advanced quantization methods for gauge theories in both the\linebreak[4]
 ${\rm Lagrangian}^{1-3}$ and ${\rm Hamiltonian}^{4-6}$ formalisms are
 based on the idea of a special type of global supersymetry, the so-called
 BRST ({\mbox Becchi} -- {\mbox Rouet} -- {\mbox Stora} -- {\mbox Tyutin})
 ${\rm symmetry.}^{7-8}$ It turns out, however, that the BRST invariance
 requirement for a theory may be strengthened by a requirement of extended
 BRST invariance. The extended BRST symmetry transformations (have been
 discussed in part in Refs.~9--12) include both the BRST and anti-BRST
 transformations (in the Yang -- Mills theories  the anti-BRST symmetry has
 been introduced in Refs.~13--14) with the fermion parameters of the BRST and
 anti-BRST transformations to form a nature doublet under the global
 symplectic group Sp(2) (see for example Ref.~12).

 The quantization rules based on the extended BRST symmetry principle for
 general gauge theories in both the Lagrangian and Hamiltonian formalisms (the
 Sp(2)-covariant quantization method) have been recently
 ${\rm proposed.}^ {15-20}$ Namely, in Refs.~15--17 an Sp(2)-covariant
 formulation of the BV ({\mbox Batalin} -- {\mbox Vilkovisky}) Lagrangian
 quantization ${\rm method}^{2,\;3}$ for general gauge theories of any stage
 reducibility has been developed. In its turn, the corresponding Hamiltonian
 version, being the Sp(2)-covariant formulation of the BFV ({\mbox Batalin} --
 {\mbox Fradkin} -- {\mbox Vilkovisky}) generalized canonical quantization
 ${\rm method}^{4-6}$ for arbitrary any-stage reducible dynamical systems with
 both first- and second-class constraints, has been suggested in
 Refs.~18--20. Note in this connection that for the first time the BFV method
 has been applied to an analysis of the extended BRST symmetry in Ref.~11.
 It should be also pointed out that in Ref.~12 in the case of dynamical systems
 subject to first-class constraints with constant structural coeffitients,
there
 has been obtained the unitarizing ${\rm Hamiltonian}^{4-6}$ invariant under
the
 extended BRST symmetry transformations for an arbitrary choice of the gauge.

 The global symplectic group Sp(2) providing a basis for the quantization
 ${\rm rules}^{15-20}$ plays the key role in the formalism proposed and
 demonstrates the advantages it yields. That is to say, in Refs.~16,~19 it is
 shown that ghost and auxiliary variables for any-stage reducible field theory
 in both the Lagrangian (fields of the total configuration spase) and
 Hamiltonian (generalized momenta and coordinates) versions form components of
 completely symmetric tensors under the group Sp(2).

 It is noteworthy that the sets of canonical variables in the framework of both
 the ${\rm standard}^{4-6}$ and Sp(2)-${\rm covariant}^{18-20}$ Hamiltonian
 formulations coincide. Meanwhile, the set of variables (specifically, the
 one of antifields) in the Sp(2)-covariant ${\rm version}^{15-17}$ of
 the Lagrangian quantization method is redundant with respect to
 the set of variables in the standard ${\rm version.}^{2,\;3}$

 The basic objects in the Hamiltonian version of the Sp(2)-covariant
 quantization method are the boson function $\cal H$ and the doublet of fermion
 functions ${\Omega^a,}^{18-20}$ while the Lagrangian version is based on the
 boson functional ${S.}^{15-17}$ The objects concerned satisfy gauge algebra
 generating ${\rm equations}^{15-20}$ and permit constructing the quantum
 action of a theory. Proof of the existence theorems for solutions to the gauge
 algebra generating equations of both the ${\rm Lagrangian}^{15-17}$ and
 ${\rm Hamiltonian}^{18-20}$ versions as well as description of arbitrariness
in
 the solutions are given in Refs.~15,~17,~18,~20.

 In this paper we shall restrict ourselves to consideration of irreducible
 gauge theories only. Namely, there are studied arbitrary dynamical systems
 with lineary independent first-class constraints in the Hamiltonian
 ${\rm version}^{18-20}$ and general gauge theories with lineary independent
 generators of gauge transformations in the Lagrangian ${\rm version.}^{15-17}$
 The explicit solutions to the generating ${\rm equations}^{15-20}$ are
obtained
 in the form of expansions in power series of ghost and auxiliary variables up
 to the third order inclusively. It is shown that in the Sp(2)-covariant
 ${\rm formulation}^{15-20}$ of both the BV and BFV quantization methods,
 solutions to the generating equations are completely defined by the gauge
 algebra structural relations as they are in the standard
 ${\rm formulation.}^{2-6}$

 Note, from the perturbation theory viewpoint, that the solutions obtained are
 quite sufficient for all practical purposes of the field theory. Indeed, an
 application of the perturbation theory implies one's knowledge of propagators
 and interaction vertices. The propagators are defined by kernels of
 differential operators present in the kinetic part of the quantum action. One
 has, therefore, apart from the initial classical action, to have at one's
 disposal the first approximation for the quantum action. The interaction
 vertices, in turn, are defined by the higher approximations (along with the
 above-mentioned ones) for solutions to the generating equations. Since the
 majority of the field theory calculations are at best the two-loop ones, the
 approximations up to the 3d order turn out to provide such calculations.
 Moreover, in the number of gauge theories with a closed algebra, the
iterations
 concerned yield, in both the {\mbox Lagrangian} and {\mbox Hamiltonian}
 versions, an exact form for the quantum action. It is these considerations
that
 made us seek solutions to the generating equations of the Sp(2)-covariant
 quantization method with an accuracy up to the 3d order in ghost and auxiliary
 variables.

 In this paper we use the condensed ${\rm notations}^{21}$ and designations of
  Refs.~15--20. The derivatives with respect to generalized momenta
 or antifields are always understood as left and those with respect to the
 corresponding  configuration space variables (i.~e. generalized coordinates
 $Q^A$ or fields $\phi^A$) as right unless specified. The left derivatives with
 respect to $Q^A$ or $\phi^A$ are labelled ``l'':
 ${\delta_{\rm l}}/{\delta Q^A}$, ${\delta_{\rm l}}/{\delta\phi^A}$. The
 {\mbox Grassmann} parity of a certain quantity $A$ is denoted $\varepsilon(A)$
 and the new ghost ${\rm number}^{15-20}$ -- ${\rm ngh}(A)$.

 We make use of the standard ${\rm definition}^{22}$ of the {\mbox Poisson}
 superbracket in an extended phase space $\Gamma=(P_A,Q^A)$
\setcounter{sct}{1}
\begin{eqnarray}
 \{F,G\}=\frac{\delta F}{\delta Q^A}\frac{\delta G}{\delta P_A} -
   \frac{\delta G}{\delta Q^A}\frac{\delta F}{\delta P_A}(-1)^
   {\varepsilon(F)\varepsilon(G)},
\end{eqnarray}
 where $P_A$ are the set of generalized momenta congugate to coordinates $Q^A$
\[ \varepsilon(P_A)=\varepsilon(Q^A),\;\;\;{\rm ngh}(P_A)=-{\rm ngh}(Q^A).\]
 The superbracket (1.1) possesses the standart algebraic properties among
 which we only point out the {\mbox Jacobi} identity
\begin{eqnarray}
 \{\{F,G\},H\}(-1)^{\varepsilon(F)\varepsilon(H)}+{\rm cycl.perm.}(FGH)
   \equiv 0.
\end{eqnarray}
\hspace*{\parindent} The indices of the global symplectic group Sp(2) are
 denoted {\it a}, {\it b}, {\it c}, \ldots, and assume two values $a=1,2$. The
 invariant tensor of the group Sp(2), being a constant antisymmetric tensor of
 rank two, is denoted $\varepsilon^{ab}$, such that $\varepsilon^{12}=1$.
 Symmetrization over the Sp(2) indices is understood in the form
\[ A^{\{ ab\}}=A^{ab}+A^{ba}.\]
{\bf 2. The Solution of Equations for ${\cal H}$, $\Omega^a$}\\[0.1mm]

 It may be convinient to remind the reader about the key points of the
 Hamiltonian version of the Sp(2)-covariant quantization
 ${\rm method.}^{18-20}$ To do this, consider the dynamical system described in
 the phase spase
\[ \eta=(p_\imath,q^\imath),\;\;\;\;{\rm ngh}(q^\imath)=0 \]
 by the classical Hamiltonian $H_0=H_0(\eta)$ and by the set of lineary
 independent first-class constraints $T_\alpha=T_\alpha(\eta)$, $\varepsilon
 (T_\alpha)\equiv\varepsilon_\alpha$ with the involution relations
\setcounter{sct}{2}
\setcounter{equation}{0}
\begin{eqnarray}
 \{T_\alpha,T_\beta\}=T_\gamma U^\gamma_{\alpha\beta},\;\;\;\;
 \{H_0,T_\alpha\}=T_\beta V^\beta_\alpha,
\end{eqnarray}
 where the structural coeffitients $U^\gamma_{\alpha\beta}$ possess the
 properties of generalized antisymmetry
\begin{eqnarray}
 U^\gamma_{\alpha\beta}=-(-1)^{\varepsilon_\alpha\varepsilon_\beta}U^\gamma_
 {\beta\alpha}.
\end{eqnarray}
 Given this, the structure of the extended phase space $\Gamma=(P_A,Q^A)$ is as
 ${\rm follows}^{18}$
\begin{eqnarray}
 \Gamma=(P_A,Q^A)=(\eta;{\cal P}_{\alpha a},C^{\alpha a};\lambda_\alpha,
 \pi^\alpha).
\end{eqnarray}
 In (2.3) $C^{\alpha a}$ form Sp(2) doublets of ghost variables
\[ \varepsilon(C^{\alpha a})=\varepsilon_\alpha+1,\;\;\;
  {\rm ngh}(C^{\alpha a})=1 \]
 and $\pi^\alpha$ are auxiliary variables
\[ \varepsilon(\pi^\alpha)=\varepsilon_\alpha,\;\;\;\;
  {\rm ngh}(\pi^\alpha)=2, \]
 introducing the gauge in the framework of the standard formulation of the
 generalized canonical quantization method.

 The boson function ${\cal H}$ and the fermion functions $\Omega^a$ introduced
 in Ref.~18 satisfy the Sp(2)-covariant generating equations
\begin{eqnarray}
 \{\Omega^a,\Omega^b\}=0,\;\;\;\;\{{\cal H},\Omega^a\}=0,
\end{eqnarray}
 with the boundary conditions
\[
   \left.\frac{\delta\Omega^a}{\delta C^{\alpha b}}\right|_{C=\pi={\cal P}
   =\lambda=0}=T_\alpha\delta^a_b,\;\;\;\;
   \left.\frac{\delta\Omega^a}{\delta\pi^\alpha}\right|_{C=\pi=\lambda=0}=
   \varepsilon^{ab}{\cal P}_{\alpha b},
\]
\begin{flushright}(2.5)\end{flushright}
\[
   \left.{\cal H}\right|_{C=\pi={\cal P}=\lambda=0}=H_0.
\]
\setcounter{equation}{5}
\hspace*{\parindent} The total unitarizing Hamiltonian $H$ is now determined
 in terms of ${\cal H}$ and $\Omega^a$ by the ${\rm formula}^{18-20}$
\begin{eqnarray}
 H={\cal H}+\frac{1}{2}\varepsilon_{ab}\{\{\Phi,\Omega^b\},\Omega^b\},
\end{eqnarray}
 where $\Phi$ is the boson function fixing a concrete choice of admissible
 gauge. An essential property of the unitarizing {\mbox Haimltonian} $H$ (2.6)
 is its invariance under the extended BRST transformations of the phase space
 $\Gamma$
\begin{eqnarray}
 \delta\Gamma=\{\Gamma,\Omega^a\}\mu_a.
\end{eqnarray}
 Here $\mu_a$ is an Sp(2) doublet of constant Grassmann parameters of extended
 BRST symmetry. Owing to the properties of the functions $\Omega^a$(2.4), the
 transformations (2.7) are nilpotent.

 For the theory in question with the Hamiltonian $H$ (2.6), the generating
 functional of the Green's functions is given in the usual form by the
 functional ${\rm integral}^{18}$
\begin{eqnarray}
 Z(I)=\int d\Gamma\;\exp\bigg\{\frac{i}{\hbar}\int dt\bigg(P_A\dot{Q}^A-H+
 I\Gamma\bigg)\bigg\}.
\end{eqnarray}
 As a consequence of invariance of the total {\mbox Hamiltonian} under the
 transformations (2.7), the vacuum functional $Z_\Phi\equiv Z(0)$ is
 ${\rm independent}^{18}$ on the choice of the gauge function $\Phi$. Indeed,
 one can readily establish that any change of the gauge $\Phi\to\Phi+\Delta
 \Phi$ in the functional integral (2.8) for $I=0$ can be compensated by the
 change of the integration variables (2.7) $\Gamma\to\Gamma+\delta\Gamma$
 with the parameters
\[
 \mu_a=\frac{i}{2\hbar}\varepsilon_{ab}\int dt\;\{\Omega^b,\Delta\Phi\}.
\]
 Hence $Z_{\Phi+\Delta\Phi}=Z_\Phi$, and therefore, the $S$ matrix is gauge
 invariant in the \mbox{Hamiltonian} version of the Sp(2)-covariant
 quantization method.

 In Ref. 18 it is shown that one can seek solutions to Eqs. (2.4) in the form
of
 expansions in power series of ghost $C^{\alpha a}$ and auxiliary $\pi^\alpha$
 variables
\begin{eqnarray}
 \Omega^a=\sum_{n=1}^\infty\Omega^a_n,\;\;\;\;{\cal H}=H_0+\sum_{n=1}^\infty
 {\cal H}_n,
\end{eqnarray}
 having required the Grassmann parity and the new ghost number to be conserved
 in every order of perturbation series
\[
 \varepsilon(\Omega^a_n)={\rm ngh}(\Omega^a_n)=1,\;\;\;\;
 \varepsilon({\cal H}_n)={\rm ngh}({\cal H}_n)=0.
\]
 In (2.9) $\Omega^a_n$ and ${\cal H}_n$ are some $n$th order polinomials in
 the variables $C^{\alpha a}$, $\pi^\alpha$. The requirement of the new ghost
 number conservation leads to the fact that $\Omega^a_n$ and ${\cal H}_n$ must
 be polinomials in ${\cal P}_{\alpha a}$, $\lambda_\alpha$ as well.

 In the first order perturbation series, the solution to Eqs. (2.4),
determining
 $\Omega^a$, can be chosen in the ${\rm form}^{18}$
\begin{eqnarray}
 \Omega^a_1=T_\alpha C^{\alpha a}+\varepsilon^{ab}{\cal P}_{\alpha
b}\pi^\alpha.
\end{eqnarray}
 Then, the higher approximations in (2.9) are determined by the
 ${\rm equations}^{18}$
\begin{eqnarray}
 W^a{\cal H}_{n+1}=D^a_{n+1},\;\;\;\;n\geq 1,
\end{eqnarray}
\begin{eqnarray}
 W^{\{a}\Omega^{b\}}_{n+1}=-B^{ab}_{n+1},\;\;\;\;n\geq 1,
\end{eqnarray}
 where the operators $W^a$ are given by the formula
\begin{eqnarray}
 W^a=T_\alpha\frac{\delta}{\delta P_{\alpha a}}+\varepsilon^{ab}{\cal P}_
 {\alpha b}\frac{\delta}{\delta\lambda_\alpha}+(-1)^{\varepsilon_\alpha}
 \varepsilon^{ab}\pi^\alpha\frac{\delta_{\rm l}}{\delta C^{\alpha b}}
\end{eqnarray}
 and possess the properties
\[ W^{\{a}W^{b\}}=0. \]
 The functions $D^a_{n+1}$ and $B^{ab}_{n+1}$ are constructed from
 $\Omega^a_{m+1}$, ${\cal H}^a_m$, $m\leq n$ by the ${\rm rules}^{18}$
\begin{eqnarray}
 B^{ab}_{n+1}=\{\Omega^a_{[n]},\Omega^b_{[n]}\}_{n+1},
\end{eqnarray}
\begin{eqnarray}
 D^a_{n+1}=\{{\cal H}_{[n]},\Omega^a_{[n+1]}\}_{n+1},
\end{eqnarray}
 where
\[
 \Omega^a_{[n]}\equiv\sum_{k=1}^n\Omega^a_k,\;\;\;\;
 {\cal H}_{[n]}\equiv H_0+\sum_{k=1}^n{\cal H}_k,
\]
 the symbol $\{,\}_k$ denotes the $k$th order for the superbracket $\{,\}$ in
 power series of the variables $C^{\alpha a}$, $\pi^\alpha$. The functions
 $B^{ab}_{n+1}$ and $D^a_{n+1}$ satisfy the equations
\[
 W^aB^{bc}_{n+1}+{\rm cycl.perm.}(abc)=0,
\]
\[
 W^{\{a}D^{b\}}_{n+1}=0,
\]
 being the compatibility conditions for Eqs. (2.11), (2.12).

 Let us give the explicit solutions to Eqs. (2.11), (2.12) when $n=1,2$ in the
 case of dynamical systems with irreducible first-class constraints described
 by the properties (2.1). The explicit form of $\Omega^a_1$ (2.10), as it
 follows from (2.14), (2.15), enables us to obtain the solutions to Eqs.
(2.11),
 (2.12) for ${\cal H}_1$, $\Omega^a_2$. Note to this end that the functions
 $D^a_1$ and $B^{ab}_2$ have the form
\begin{eqnarray}
 D^a_1=\{H_0,T_\alpha\}C^{\alpha a},
\end{eqnarray}
\begin{eqnarray}
 B^{ab}_2=-\{T_\beta,T_\alpha\}(-1)^{\varepsilon_\beta}C^{\alpha a}C^{\beta b}.
\end{eqnarray}
 It is convinient, seeking the solution to Eqs. (2.12) with the right-hand side
 (2.17), to make use of the following decomposition of an Sp(2) tensor of rank
 two, constructed from $C^{\alpha a}$
\begin{eqnarray}
 (-1)^{\varepsilon_\beta}C^{\alpha a}C^{\beta b}=T^{\alpha\beta\{ab\}}+
 \hat{T}^{\alpha\beta}\varepsilon^{ab},
\end{eqnarray}
 where the components of decomposition
\[
 T^{\alpha\beta\{ab\}}=\frac{1}{2}(-1)^{\varepsilon_\beta}C^{\alpha\{a}
 C^{\beta b\}},\;\;\;\;\hat{T}^{\alpha\beta}\varepsilon^{ab}=-\frac{1}{2}(-1)^
 {\varepsilon_\beta}\varepsilon_{cd}C^{\alpha c}C^{\beta d}\varepsilon^{ab}
\]
 form respectively the symmetric and antisymmetric tensors of rank two under
 the group Sp(2). The quantities $T^{\alpha\beta\{ab\}}$ and
 $\hat{T}^{\alpha\beta}$ possess the following properties of generalized
 (anti)symmetry
\begin{eqnarray}
 T^{\alpha\beta\{ab\}}=-(-1)^{\varepsilon_\alpha\varepsilon_\beta}T^
 {\beta\alpha\{ab\}},\;\;\;\;
 \hat{T}^{\alpha\beta}=(-1)^{\varepsilon_\alpha\varepsilon_\beta}\hat{T}^
 {\beta\alpha}.
\end{eqnarray}
 Then, with allowance made for the involution relations (2.1), we find from the
 definition (2.18) and the properties (2.2), (2.19) the following
 representations for $D^a_1$, $B^{ab}_2$
\begin{eqnarray}
 D^a_1=T_\beta V^\beta_\alpha C^{\alpha a},
\end{eqnarray}
\begin{eqnarray}
 B^{ab}_2=-T_\gamma U^\gamma_{\beta\alpha}T^{\alpha\beta\{ab\}}.
\end{eqnarray}
\hspace*{\parindent} Solving Eqs. (2.11) with the right-hand side (2.20) is as
 follows. It is necessary to choose such a function $X_1$ as to produce all (or
 several) structures present in $D^a_1$ when the operators $W^a$ act upon it
\[ X_1={\cal P}_{\beta a}V^\beta_\alpha C^{\alpha
 a}.
\]
 Then, applying $W^a$ to $X_1$
\begin{eqnarray} W^aX_1=T_\beta
 V^\beta_\alpha C^{\alpha a}-{\cal P}_{\beta b}V^\beta_\alpha
 \pi^\alpha\varepsilon^{ab},
\end{eqnarray}
 we find that the first summand in the right-hand side (2.22) and $D^a_1$
 (2.20) coincide. One readily observes that the second summand in (2.22) is
 reproduced by applying $W^a$ to the function $X_2$ of the form
\[
 X_2=\lambda_\beta V^\beta_\alpha\pi^\alpha,
\]
 namely
\[
  W^aX_2={\cal P}_{\beta b}V^\beta_\alpha\pi^\alpha\varepsilon^{ab}.
\]
 Given this, it follows from the obvious relation
\[
  W^a(X_1+X_2)-D^a_1=0
\]
 that one can choose the solution to Eqs. (2.11) with the right-hand side
 (2.20) in the form
\begin{eqnarray}
 {\cal H}_1=X_1+X_2={\cal P}_{\beta a}V^\beta_\alpha C^{\alpha a}+\lambda_\beta
 V^\beta_\alpha\pi^\alpha.
\end{eqnarray}
\hspace*{\parindent} The method of solving Eqs. (2.12) with the right-hand
 side (2.21) is quite analogus to the one considered above. To this end, it
 suffices to apply the operators $W^a$ to the functions of the form
\[
  X^a_1=\frac{1}{2}{\cal P}_{\gamma b}U^\gamma_{\beta\alpha}T^
  {\alpha\beta\{ab\}},\;\;\;\;X^a_2=-\frac{1}{2}\lambda_\gamma U^\gamma_
  {\beta\alpha}(-1)^{\varepsilon_\beta}C^{\alpha a}\pi^\beta,
\]
 then
\[ W^{\{a}X^{b\}}_1=T_\gamma U^\gamma_{\beta\alpha}T^{\alpha\beta\{ab\}}+\frac
   {1}{2}{\cal P}_{\gamma c}U^\gamma_{\beta\alpha}(-1)^{\varepsilon_\beta}
   C^{\alpha\{a}\varepsilon^{b\}c}\pi^\beta,
\]
\begin{eqnarray}
 W^{\{a}X^{b\}}_2=-\frac{1}{2}{\cal P}_{\gamma c}U^\gamma_{\beta\alpha}(-1)^
 {\varepsilon_\beta}C^{\alpha\{a}\varepsilon^{b\}c}\pi^\beta.
\end{eqnarray}
 In (2.24) we made allowance for the generalized antisymmetry property (2.2) of
 the gauge algebra structural coeffitients $U^\gamma_{\alpha\beta}$, and, as a
 consequence, for the identity
\[
 U^\gamma_{\alpha\beta}\pi^\beta\pi^\alpha\equiv 0.
\]
 Then, making a comparison between $B^{ab}_2$ and $W^{\{a}X^{b\}}_1$,
 $W^{\{a}X^{b\}}_2$, we have
\[
  W^{\{a}X^{b\}}_1+W^{\{a}X^{b\}}_2+B^{ab}_2=0.
\]
 Consequently, the solution to Eqs. (2.12) with the right-hand side (2.21) can
 be chosen in the form
\begin{eqnarray}
 \Omega^a_2=X^a_1+X^a_2=\frac{1}{2}{\cal P}_{\gamma b}U^\gamma_{\beta\alpha}T^
 {\alpha\beta\{ab\}}-\frac{1}{2}\lambda_\gamma U^\gamma_{\beta\alpha}(-1)^
 {\varepsilon_\beta}C^{\alpha a}\pi^\beta.
\end{eqnarray}
\hspace*{\parindent} Now consider, taking into account the explicit form of the
 functions ${\cal H}_{[1]}$ (2.23) and $\Omega^a_{[2]}$ (2.10), (2.25), the
 solution of Eqs. (2.11), (2.12) for ${\cal H}_2$, $\Omega^a_3$. By virtue of
 (2.14), (2.15), we find $D^a_2$ and $B^{ab}_3$ in the form
\begin{eqnarray}
 D^a_2&=&\frac{1}{2}{\cal P}_{\gamma b}\bigg(\{H_0,U^\gamma_{\beta\alpha}\}+
 V^\gamma_\delta U^\delta_{\beta\alpha}-U^\gamma_{\beta\delta}V^\delta_\alpha+
 (-1)^{\varepsilon_\alpha\varepsilon_\beta}U^\gamma_{\alpha\delta}V^\delta_
 \beta-\{V^\gamma_\beta,T_\alpha\}\nonumber\\&&
 +(-1)^{\varepsilon_\alpha\varepsilon_\beta}
 \{V^\gamma_\alpha,T_\beta\}\bigg)T^{\alpha\beta\{ab\}}-{\cal P}_{\gamma b}
 \{V^\gamma_\beta,T_\alpha\}\hat{T}^{\alpha\beta}\varepsilon^{ab}\nonumber\\&&
 -\frac{1}{2}\lambda_\gamma\bigg(\{H_0,U^\gamma_{\beta\alpha}\}+V^\gamma_\delta
 U^\delta_{\beta\alpha}-U^\gamma_{\beta\delta}V^\delta_\alpha+(-1)^
 {\varepsilon_\alpha\varepsilon_\beta}U^\gamma_{\alpha\delta}V^\delta_\beta
 \nonumber\\&&
 -2\{V^\gamma_\beta,T_\alpha\}\bigg)(-1)^{\varepsilon_\beta}C^{\alpha a}\pi^
 \beta,\nonumber\\&&\\
 B^{ab}_3&=&-\frac{1}{2}{\cal P}_{\rho c}\bigg(\frac{1}{2}U^\rho_
 {\gamma\delta}U^\delta_{\beta\alpha}(-1)^{\varepsilon_\alpha\varepsilon_
 \gamma}+\frac{1}{2}U^\rho_{\alpha\delta}U^\delta_{\gamma\beta}(-1)^
 {\varepsilon_\alpha\varepsilon_\beta}\nonumber\\&&
 -\{U^\rho_{\gamma\beta},T_\alpha\}(-1)^{\varepsilon_\alpha\varepsilon_\gamma}
 \bigg)(-1)^{\varepsilon_\beta+\varepsilon_\alpha\varepsilon_\gamma}
 C^{\alpha\{a}C^{\beta b\}}C^{\gamma c}\nonumber\\&&
 -\frac{1}{2}\lambda_\rho\bigg(
 \frac{1}{2}U^\rho_{\gamma\delta}U^\delta_{\beta\alpha}(-1)^{\varepsilon_\alpha
 \varepsilon_\gamma}+\frac{1}{2}U^\rho_{\alpha\delta}U^\delta_{\gamma\beta}
 (-1)^{\varepsilon_\alpha\varepsilon_\beta}\nonumber\\&&
 -\{U^\rho_{\gamma\beta},T_\alpha\}(-1)^{\varepsilon_\alpha\varepsilon_\gamma}
 \bigg)(-1)^{\varepsilon_\alpha\varepsilon_\gamma}T^{\alpha\beta\{ab\}}\pi^
 \gamma,\nonumber
\end{eqnarray}
 here we made use of the definition (2.18) and the properties (2.19).

 Let us now introduce, for the sake of convinience, the following decomposition
 of an Sp(2) tensor of rank three present in $B^{ab}_3$
\begin{eqnarray}
 \frac{1}{2}(-1)^{\varepsilon_\beta+\varepsilon_\alpha\varepsilon_\beta}
 C^{\alpha a}C^{\beta b}C^{\gamma c}&=&\frac{1}{2}T^{\alpha\beta\gamma\{abc\}}
 \nonumber\\&&
 +\frac{1}{3}\varepsilon^{ab}\bigg(\hat{T}^{\alpha\beta}C^{\gamma c}(-1)^
 {\varepsilon_\alpha\varepsilon_\gamma}-\hat{T}^{\beta\gamma}C^{\alpha c}(-1)^
 {\varepsilon_\alpha\varepsilon_\beta}\bigg)\nonumber\\&&
 +\frac{1}{3}\varepsilon^{ac}\bigg(\hat{T}^{\beta\gamma}C^{\alpha b}(-1)^
 {\varepsilon_\alpha\varepsilon_\beta}-\hat{T}^{\gamma\alpha}C^{\beta b}(-1)^
 {\varepsilon_\beta\varepsilon_\gamma}\bigg),\nonumber\\
\end{eqnarray}
 where
\begin{eqnarray}
 T^{\alpha\beta\gamma\{abc\}}=\frac{1}{3}(-1)^{\varepsilon_\alpha\varepsilon_
 \gamma}T^{\alpha\beta\{ab\}}C^{\gamma c}+{\rm cycl.perm.}(abc)
\end{eqnarray}
 is a symmetric tensor of rank three under the group Sp(2). From the definition
 of the tensor (2.28) there follow the properties
\begin{eqnarray}
 T^{\alpha\beta\gamma\{abc\}}=-(-1)^{\varepsilon_{\alpha\beta\gamma}}T^{\beta
 \alpha\gamma\{abc\}}=-(-1)^{\varepsilon_{\alpha\beta\gamma}}T^{\alpha\gamma
 \beta\{abc\}},
\end{eqnarray}
 where
\[
 \varepsilon_{\alpha\beta\gamma}\equiv\varepsilon_\alpha\varepsilon_\beta+
 \varepsilon_\alpha\varepsilon_\gamma+\varepsilon_\beta\varepsilon_\gamma.
\]
\hspace*{\parindent} To solve Eqs. (2.11), (2.12) for ${\cal H}_2$,
 $\Omega^a_3$ it is necessary to make use of the gauge algebra structural
 relations, being the consequence of the involution relations (2.1), namely
\begin{eqnarray}
 T_\delta\bigg(\{H_0,U^\delta_{\alpha\beta}\} + V^\delta_\gamma U^\gamma_
 {\alpha\beta} - U^\delta_{\alpha\gamma}V^\gamma_\beta
 + (-1)^{\varepsilon_\alpha\varepsilon_\beta}U^\delta_{\beta\gamma}
 V^\gamma_\alpha\nonumber\\
 - \{V^\delta_\alpha,T_\beta\} + (-1)^{\varepsilon_\alpha\varepsilon_\beta}
 \{V^\delta_\beta,T_\alpha\}\bigg)\equiv 0,
\end{eqnarray}
\begin{eqnarray}
 T_\rho\bigg(\{U^\rho_{\alpha\beta},T_\gamma\}-U^\rho_{\alpha\delta}U^\delta_
 {\beta\gamma}\bigg)(-1)^{\varepsilon_\alpha\varepsilon_\gamma}+
 {\rm cycl.perm.}(\alpha\beta\gamma)\equiv 0.
\end{eqnarray}
 The relations (2.30), (2.31) follow respectively from the {\mbox Jacobi}
 identities (1.2) of the form
\[
 \{\{T_\alpha,T_\beta\},H_0\}+\{\{H_0,T_\alpha\},T_\beta\}+\{\{T_\beta,H_0\},
 T_\alpha\}(-1)^{\varepsilon_\alpha\varepsilon_\beta}\equiv 0,
\]
\[
 \{\{T_\alpha,T_\beta\},T_\gamma\}(-1)^{\varepsilon_\alpha\varepsilon_\gamma}+
 {\rm cycl.perm.}(\alpha\beta\gamma)\equiv 0.
\]
 Then, by virtue of lineary independence of the constraints $T_\alpha$, we
 conclude that there exist such structural coeffitients $E^{\gamma\delta}_
 {\alpha\beta}$ and $E^{\delta\rho}_{\alpha\beta\gamma}$ with the properties
\[
 E^{\gamma\delta}_{\alpha\beta}=-(-1)^{\varepsilon_\alpha\varepsilon_\beta}E^
 {\gamma\delta}_{\beta\alpha}=-(-1)^{\varepsilon_\gamma\varepsilon_\delta}E^
 {\delta\gamma}_{\alpha\beta},
\]
\begin{flushright}(2.32)\end{flushright}
\setcounter{equation}{32}
\[
 E^{\delta\rho}_{\alpha\beta\gamma}=-(-1)^{\varepsilon_\rho\varepsilon_\delta}
 E^{\rho\delta}_{\alpha\beta\gamma},\;\;\;\;E^{\delta\rho}_{\alpha\beta\gamma}=
 E^{\delta\rho}_{\gamma\alpha\beta},
\]
 that the relations (2.30), (2.31) could be represented in the form
\begin{eqnarray}
 T_\gamma E^{\gamma\delta}_{\alpha\beta}&=&\frac{1}{2}\bigg(\{H_0,U^\delta_
 {\alpha\beta}\} + V^\delta_\gamma U^\gamma_{\alpha\beta} - U^\delta_
 {\alpha\gamma}V^\gamma_\beta + (-1)^{\varepsilon_\alpha\varepsilon_\beta}U^
 \delta_{\beta\gamma}V^\gamma_\alpha\nonumber\\&&
 - \{V^\delta_\alpha,T_\beta\} + (-1)^
 {\varepsilon_\alpha\varepsilon_\beta}\{V^\delta_\beta,T_\alpha\}\bigg),
\end{eqnarray}
\begin{eqnarray}
 T_\delta E^{\delta\rho}_{\alpha\beta\gamma}=\frac{1}{3}\bigg(U^\rho_
 {\alpha\delta}U^\delta_{\beta\gamma}-\{U^\rho_{\alpha\beta},T_\gamma\}\bigg)
 (-1)^{\varepsilon_\alpha\varepsilon_\gamma}+{\rm cycl.perm.}
 (\alpha\beta\gamma).
\end{eqnarray}
 From (2.33), (2.34) with allowance made for the properties (2.29), there
 follow the representations for the functions $D^a_2$ and $B^{ab}_3$
\begin{eqnarray}
 D^a_2&=&{\cal P}_{\delta b}T_\rho E^{\rho\delta}_{\beta\alpha}T^{\alpha\beta
 \{ab\}} - {\cal P}_{\gamma b}\{V^\gamma_\beta,T_\alpha\}\hat{T}^{\alpha\beta}
 \varepsilon^{ab}\nonumber\\&&
 - \frac{1}{2}\lambda_\gamma\bigg(\{H_0,U^\gamma_
 {\beta\alpha}\} + V^\gamma_\delta U^\delta_{\beta\alpha} - U^\gamma_
 {\beta\delta}V^\delta_\alpha + (-1)^{\varepsilon_\alpha\varepsilon_\beta}U^
 \gamma_{\alpha\delta}V^\delta_\beta\nonumber\\&&
 - 2\{V^\gamma_\beta,T_\alpha\}\bigg)(-1)^
 {\varepsilon_\beta}C^{\alpha a}\pi^\beta,
\end{eqnarray}
\begin{eqnarray}
 B^{ab}_3&=&-{\cal P}_{\delta c}T_\rho E^{\rho\delta}_{\gamma\beta\alpha}T^
 {\alpha\beta\gamma\{abc\}} - \frac{1}{3}{\cal P}_{\rho c}\bigg(\{U^\rho_
 {\beta\alpha},T_\gamma\}(-1)^{\varepsilon_\gamma(\varepsilon_\alpha +
 \varepsilon_\beta)}\nonumber\\&&
 + \frac{1}{2}U^\rho_{\gamma\delta}U^\delta_{\beta\alpha}
 \bigg)(-1)^{\varepsilon_\beta+\varepsilon_\gamma}C^{\alpha\{a}
 \varepsilon^{b\}c}\hat{T}^{\beta\gamma}\nonumber\\&&
 - \lambda_\rho\bigg(\frac{1}{2}(-1)^{\varepsilon_\alpha
 \varepsilon_\gamma}U^\rho_{\gamma\delta}U^\delta_{\beta\alpha} + \frac{1}{2}
 (-1)^{\varepsilon_\alpha\varepsilon_\beta}U^\rho_{\alpha\delta}U^\delta_
 {\gamma\beta}\nonumber\\&&
 - (-1)^{\varepsilon_\alpha\varepsilon_\gamma}\{U^\rho_
 {\gamma\beta},T_\alpha\}\bigg)(-1)^{\varepsilon_\alpha\varepsilon_\gamma}T^
 {\alpha\beta\{ab\}}\pi^\gamma,
\end{eqnarray}
 To solve Eqs. (2.11) with the right-hand side (2.35) it is necessary to take
 into account the structural relations (2.33) and the identity
\[
 E^{\gamma\delta}_{\alpha\beta}\pi^\beta\pi^\alpha\equiv 0,
\]
 which follows from the properties of generalized antisymmetry (2.32) for the
 structural coeffitients $E^{\gamma\delta}_{\alpha\beta}$. In turn, solving
 Eqs. (2.12) with the right-hand side (2.36) involves making use of the
 relations (2.34) with allowance made for the obvious identity
\[
 \bigg(E^{\delta\rho}_{\alpha\beta\gamma}(-1)^{\varepsilon_\alpha\varepsilon_
 \gamma} -
(-1)^{\varepsilon_\alpha\varepsilon_\beta}E^{\delta\rho}_{\beta\alpha
 \gamma}(-1)^{\varepsilon_\beta\varepsilon_\gamma}\bigg)\pi^\beta\pi^\alpha
 \equiv 0.
\]
 It is noteworthy that the methods of solving the equations for ${\cal H}_2$
 and $\Omega^a_3$ are quite similar to each other.

 Now, turnig ourselves to solution of the equations for ${\cal H}_2$, note
 that the structures present in $D^a_2$
\[
 {\cal P}_{\gamma b}T_\delta E^{\delta\gamma}_{\beta\alpha}T^
 {\alpha\beta\{ab\}},\;\;\;\;{\cal P}_{\gamma b}\{V^\gamma_\beta,T_\alpha\}
 \hat{T}^{\alpha\beta}\varepsilon^{ab}
\]
 could be reproduced by applying the operators $W^a$ respectively to the
 functions
\[
 \frac{1}{2}{\cal P}_{\gamma a}{\cal P}_{\delta b}(-1)^{\varepsilon_\gamma}E^
 {\delta\gamma}_{\beta\alpha}T^{\alpha\beta\{ab\}},\;\;\;\;
 \lambda_\gamma\{V^\gamma_\beta,T_\alpha\}\hat{T}^{\alpha\beta},
\]
 namely
\begin{eqnarray}
 W^a\bigg(-\frac{1}{2}{\cal P}_{\gamma b}{\cal P}_{\delta c}(-1)^{\varepsilon_
 \gamma}E^{\delta\gamma}_{\beta\alpha}T^{\alpha\beta\{bc\}}\bigg)&=&
 {\cal P}_{\gamma b}T_\delta E^{\delta\gamma}_{\beta\alpha}T^
 {\alpha\beta\{ab\}}\nonumber\\&&
 + {\cal P}_{\gamma b}{\cal P}_{\delta c}(-1)^{\varepsilon_\gamma}E^
 {\delta\gamma}_{\beta\alpha}(-1)^{\varepsilon_\beta}C^{\alpha c}\pi^\beta
 \varepsilon^{ab},\nonumber
\end{eqnarray}
\begin{eqnarray}
 W^a\bigg(\lambda_\gamma\{V^\gamma_\beta,T_\alpha\}\hat{T}^{\alpha\beta}\bigg)
 &=&{\cal P}_{\gamma b}\{V^\gamma_\beta,T_\alpha\}\hat{T}^{\alpha\beta}
 \varepsilon^{ab}\nonumber\\&&
 -\frac{1}{2}\lambda_\gamma\bigg(\{V^\gamma_\beta,T_\alpha\} + (-1)^
 {\varepsilon_\alpha\varepsilon_\beta}\{V^\gamma_\alpha,T_\beta\}\bigg)(-1)^
 {\varepsilon_\beta}C^{\alpha a}\pi^\beta.\nonumber
\end{eqnarray}
 Consider the function $Y_1$
\[
 Y_1=-\frac{1}{2}{\cal P}_{\gamma a}{\cal P}_{\delta b}(-1)^
 {\varepsilon_\gamma}E^{\delta\gamma}_{\beta\alpha}T^{\alpha\beta\{ab\}}-
 \lambda_\gamma\{V^\gamma_\beta,T_\alpha\}\hat{T}^{\alpha\beta},
\]
 then from comparison of $D^a_2$ with $W^aY_1$ we find
\begin{eqnarray}
 W^aY_1-D^a_2&=&{\cal P}_{\gamma b}{\cal P}_{\delta c}(-1)^{\varepsilon_
 \gamma}E^{\delta\gamma}_{\beta\alpha}(-1)^{\varepsilon_\beta}C^{\alpha c}\pi^
 \beta\varepsilon^{ab}\nonumber\\&&
 +\lambda_\gamma T_\delta E^{\delta\gamma}_{\beta\alpha}(-1)^
 {\varepsilon_\beta}C^{\alpha a}\pi^\beta.
\end{eqnarray}
 It is clear from analysis of summands in the right-hand side (2.37) that the
 structures present there could be reproduced by applying the operators $W^a$
 to the function $Y_2$ of the form
\[
 Y_2=-\lambda_\gamma{\cal P}_{\delta a}(-1)^{\varepsilon_\gamma}E^
 {\delta\gamma}_{\beta\alpha}(-1)^{\varepsilon_\beta}C^{\alpha a}\pi^\beta.
\]
 Indeed,
\[
 W^aY_2=-{\cal P}_{\gamma b}{\cal P}_{\delta c}(-1)^{\varepsilon_\gamma}E^
 {\delta\gamma}_{\beta\alpha}(-1)^{\varepsilon_\beta}C^{\alpha c}\pi^\beta
 \varepsilon^{ab}-\lambda_\gamma T_\delta E^{\delta\gamma}_{\beta\alpha}(-1)^
 {\varepsilon_\beta}C^{\alpha a}\pi^\beta
\]
 and we find that
\[
  W^a\bigg(Y_1+Y_2\bigg)-D^a_2=0.
\]
 Consequently, the solution to Eqs. (2.11) with the right-hand side (2.35) has
 the form
\begin{eqnarray}
 {\cal H}_2\;=\;Y_1+Y_2&=&-\frac{1}{2}{\cal P}_{\gamma a}{\cal P}_{\delta b}
 (-1)^{\varepsilon_\gamma}E^{\delta\gamma}_{\beta\alpha}T^{\alpha\beta\{ab\}}
 \nonumber\\&&
 -\lambda_\gamma{\cal P}_{\delta a}(-1)^{\varepsilon_\gamma}E^{\delta\gamma}
 _{\beta\alpha}(-1)^{\varepsilon_\beta}C^{\alpha a}\pi^\beta
 -\lambda_\gamma\{V^\gamma_\beta,T_\alpha\}\hat{T}^{\alpha\beta}.\nonumber\\
\end{eqnarray}
\hspace*{\parindent} Omitting details of calculation of the functions
 $\Omega^a_3$, we only give here their resultant form
\begin{eqnarray}
 \Omega^a_3&=&-\frac{1}{4}{\cal P}_{\delta b}{\cal P}_{\rho c}(-1)^
 {\varepsilon_\delta}E^{\rho\delta}_{\gamma\beta\alpha}T^
 {\alpha\beta\gamma\{abc\}}+\frac{1}{2}\lambda_\delta{\cal P}_{\rho b}(-1)^
 {\varepsilon_\delta}E^{\rho\delta}_{\gamma\beta\alpha}(-1)^
 {\varepsilon_\alpha\varepsilon_\gamma}T^{\alpha\beta\{ab\}}\pi^\gamma
 \nonumber\\&&
 +\frac{1}{3}\lambda_\rho\bigg(\frac{1}{2}U^\rho_{\gamma\delta}U^\delta_
 {\beta\alpha}+(-1)^{\varepsilon_\gamma(\varepsilon_\alpha+\varepsilon_\beta)}
 \{U^\rho_{\beta\alpha},T_\gamma\}\bigg)(-1)^{\varepsilon_\beta+\varepsilon_
 \gamma}C^{\alpha a}\hat{T}^{\beta\gamma}.
\end{eqnarray}
\hspace*{\parindent} The explicit form of the functions ${\cal H}_{[2]}$
 (2.23), (2.38) and $\Omega^a_{[3]}$ (2.10), (2.25), (2.39) obtained above
 enables us to consider the solution of Eqs. (2.11) for ${\cal H}_3$. To this
 end, it is necessary, as before, to turn ourselves to subsequent gauge algebra
 structural relations, being of the form
\begin{eqnarray}
 T_\delta E^{\delta\rho\sigma}_{\alpha\beta\gamma}&=&-\frac{1}{12}\{H_0,E^
 {\rho\sigma}_{\alpha\beta\gamma}\}-\frac{1}{12}\bigg(V^\rho_\delta E^{\delta
 \sigma}_{\alpha\beta\gamma}-(-1)^{\varepsilon_\rho\varepsilon_\sigma}V^\sigma
 _\delta E^{\delta\rho}_{\alpha\beta\gamma}\bigg)\nonumber\\&&
 +\frac{1}{4}E^{\rho\sigma}_{\alpha\beta\delta}V^\delta_\gamma(-1)^
 {\varepsilon_\alpha(\varepsilon_\gamma+\varepsilon_\delta)}+\frac{1}{6}(-1)^
 {\varepsilon_\alpha\varepsilon_\gamma}\bigg(E^{\rho\sigma}_{\alpha\delta}U^
 \delta_{\beta\gamma}-\{E^{\rho\sigma}_{\alpha\beta},T_\gamma\}\bigg)
 \nonumber\\&&
 +\frac{1}{12}(-1)^{\varepsilon_\alpha\varepsilon_\gamma}\bigg(\{V^\rho_\alpha,
 U^\sigma_{\beta\gamma}\}(-1)^{\varepsilon_\alpha\varepsilon_\sigma}-(-1)^
 {\varepsilon_\rho\varepsilon_\sigma}\{V^\sigma_\alpha,U^\rho_{\beta\gamma}\}
 (-1)^{\varepsilon_\alpha\varepsilon_\rho}\bigg)\nonumber\\&&
 +\frac{1}{6}(-1)^{\varepsilon_\alpha\varepsilon_\gamma}\bigg(U^\rho_
 {\alpha\delta}E^{\delta\sigma}_{\beta\gamma}(-1)^{\varepsilon_\alpha
 \varepsilon_\sigma}-(-1)^{\varepsilon_\rho\varepsilon_\sigma}U^\sigma_{\alpha
 \delta}E^{\delta\rho}_{\beta\gamma}(-1)^{\varepsilon_\alpha\varepsilon_\rho}
 \bigg)\nonumber\\&&
 +{\rm cycl.perm.}(\alpha\beta\gamma),
\end{eqnarray}
 where the structural coeffitients $E^{\delta\rho\sigma}_{\alpha\beta\gamma}$
 possess the properties
\begin{eqnarray}
 E^{\delta\rho\sigma}_{\alpha\beta\gamma}=-(-1)^{\varepsilon_\delta\varepsilon_
 \rho}E^{\rho\delta\sigma}_{\alpha\beta\gamma}=-(-1)^{\varepsilon_\rho
 \varepsilon_\sigma}E^{\delta\sigma\rho}_{\alpha\beta\gamma},\;\;\;\;
 E^{\delta\rho\sigma}_{\alpha\beta\gamma}=E^{\delta\rho\sigma}_{\gamma\alpha
 \beta}.
\end{eqnarray}
 Validity of (2.40), (2.41) follows from the structural relations (2.33),
(2.34)
 with allowance made for the {\mbox Jacobi} identities
\begin{eqnarray}
 \{\{T_\alpha,T_\beta\},V^\delta_\gamma\}(-1)^{\varepsilon_\alpha(\varepsilon
 _\gamma+\varepsilon_\delta)}+\{\{T_\beta,V^\delta_\gamma\},T_\alpha\}(-1)^
 {\varepsilon_\alpha\varepsilon_\beta}\nonumber\\
 +\{\{V^\delta_\gamma,T_\alpha\},T_\beta\}(-1)^{\varepsilon_\beta(\varepsilon_
 \gamma+\varepsilon_\delta)}\equiv 0,\nonumber
\end{eqnarray}
\begin{eqnarray}
 \{\{T_\alpha,U^\delta_{\beta\gamma}\},H_0\}+\{\{H_0,T_\alpha\},U^\delta_
 {\beta\gamma}\}+\{\{U^\delta_{\beta\gamma},H_0\},T_\alpha\}(-1)^{\varepsilon_
 \alpha(\varepsilon_\beta+\varepsilon_\gamma+\varepsilon_\delta)}\equiv 0,
 \nonumber
\end{eqnarray}
 which could be presented respectively in the form
\begin{eqnarray}
 T_\delta\{V^\rho_\alpha,U^\delta_{\beta\gamma}\}(-1)^{\varepsilon_\delta
 (\varepsilon_\alpha+\varepsilon_\rho)}&=&\bigg(\{\{V^\rho_\alpha,T_\beta\},T_
 \gamma\}-(-1)^{\varepsilon_\beta\varepsilon_\gamma}\{\{V^\rho_\alpha,
 T_\gamma\},T_\beta\}\bigg)\nonumber\\&&
 -\{V^\rho_\alpha,T_\delta\}U^\delta_{\beta\gamma},\nonumber
\end{eqnarray}
\begin{flushright}(2.42)\end{flushright}
\setcounter{equation}{42}
\begin{eqnarray}
 T_\delta\{V^\delta_\alpha,U^\rho_{\beta\gamma}\}(-1)^{\varepsilon_\alpha(
 \varepsilon_\beta+\varepsilon_\gamma+\varepsilon_\rho)}&=&\{U^\rho_{\beta
 \gamma},T_\delta\}V^\delta_\alpha+\{\{U^\rho_{\beta\gamma},T_\alpha\},H_0\}
 \nonumber\\&&
 +\{\{H_0,U^\rho_{\beta\gamma}\},T_\alpha\},\nonumber
\end{eqnarray}
 Then, taking into account the definition (2.27) and the properties (2.29), we
 obtain the functions $D^a_3$ in the form
\begin{eqnarray*}
 D^a_3&=&{\cal P}_{\delta b}{\cal P}_{\rho c}(-1)^{\varepsilon_\delta}T_\sigma
 E^{\sigma\rho\delta}_{\gamma\beta\alpha}T^{\alpha\beta\gamma\{abc\}}-
 \frac{1}{3}{\cal P}_{\rho b}{\cal P}_{\sigma c}(-1)^{\varepsilon_\rho}\bigg[
 2U^\sigma_{\gamma\delta}E^{\delta\rho}_{\beta\alpha}(-1)^{\varepsilon_\gamma
 \varepsilon_\rho}\\&& 
  +U^\rho_{\gamma\delta}E^{\delta\sigma}_{\beta\alpha}(-1)^{\varepsilon_\sigma(
 \varepsilon_\gamma+\varepsilon_\rho)}-E^{\sigma\rho}_{\gamma\delta}U^\delta_
 {\beta\alpha}-2\{E^{\sigma\rho}_{\gamma\alpha},T_\beta\}(-1)^{\varepsilon_
 \alpha\varepsilon_\beta}\\&& 
 +\{V^\sigma_\gamma,U^\rho_{\beta\alpha}\}(-1)^{\varepsilon_\gamma\varepsilon_
 \rho}+2\{V^\rho_\gamma,U^\sigma_{\beta\alpha}\}(-1)^{\varepsilon_\sigma(
 \varepsilon_\gamma+\varepsilon_\rho)}\bigg](-1)^{\varepsilon_\beta+
 \varepsilon_\gamma}\varepsilon^{ab}C^{\alpha c}\hat{T}^{\beta\gamma}\\&& 
 +\lambda_\rho{\cal P}_{\sigma b}(-1)^{\varepsilon_\rho}\bigg[\frac{1}{2}\bigg(
 \{H_0,E^{\sigma\rho}_{\gamma\beta\alpha}\}+V^\sigma_\delta E^{\delta\rho}_
 {\gamma\beta\alpha}-(-1)^{\varepsilon_\rho\varepsilon_\sigma}V^\rho_\delta E^
 {\delta\sigma}_{\gamma\beta\alpha}\bigg)(-1)^{\varepsilon_\alpha\varepsilon_
 \gamma}\\&& 
  -\frac{1}{2}\bigg(E^{\sigma\rho}_{\gamma\beta\delta}V^\delta_\alpha(-1)^
 {\varepsilon_\gamma(\varepsilon_\alpha+\varepsilon_\delta)}+{\rm cycl.perm}
 (\alpha\beta\gamma)\bigg)(-1)^{\varepsilon_\alpha\varepsilon_\gamma}\\&& 
  +\frac{1}{2}\{V^\rho_\gamma,U^\sigma_{\beta\alpha}\}(-1)^{\varepsilon_\sigma(
 \varepsilon_\gamma+\varepsilon_\rho)}+\frac{1}{2}\{V^\sigma_\beta,U^\rho_
 {\gamma\alpha}\}(-1)^{\varepsilon_\beta(\varepsilon_\gamma+\varepsilon_\rho)}+
 \{E^{\sigma\rho}_{\gamma\beta},T_\alpha\}\\&& 
  -\frac{1}{2}\bigg(E^{\sigma\rho}_{\gamma\delta}U^\delta_{\beta\alpha}-(-1)^
 {\varepsilon_\beta\varepsilon_\gamma}E^{\sigma\rho}_{\beta\delta}U^\delta_
 {\gamma\alpha}\bigg)+\frac{1}{2}\bigg(U^\rho_{\gamma\delta}E^{\delta\sigma}_
 {\beta\alpha}(-1)^{\varepsilon_\sigma(\varepsilon_\gamma+\varepsilon_\rho)}
 \\&& 
  -(-1)^{\varepsilon_\gamma\varepsilon_\beta}U^\rho_{\beta\delta}E^
 {\delta\sigma}_{\gamma\alpha}(-1)^{\varepsilon_\sigma(\varepsilon_\beta+
 \varepsilon_\rho)}\bigg)+U^\sigma_{\beta\delta}E^{\delta\rho}_{\gamma\alpha}
 (-1)^{\varepsilon_\beta(\varepsilon_\gamma+\varepsilon_\rho)}\bigg]T^
 {\alpha\beta\{ab\}}\pi^\gamma\\&& 
  +\lambda_\rho{\cal P}_{\sigma
b}(-1)^{\varepsilon_\rho}\bigg[\{E^{\sigma\rho}_
 {\gamma\beta},T_\alpha\}+\frac{1}{2}\{V^\sigma_\beta,U^\rho_{\gamma\alpha}\}
 (-1)^{\varepsilon_\beta(\varepsilon_\rho+\varepsilon_\gamma)}\\&& 
 +\frac{1}{2}U^\rho_{\beta\delta}E^{\delta\sigma}_{\gamma\alpha}(-1)^
 {\varepsilon_\rho\varepsilon_\sigma+\varepsilon_\beta(\varepsilon_\gamma+
 \varepsilon_\sigma)}+\frac{1}{2}(-1)^{\varepsilon_\beta\varepsilon_\gamma}E^
 {\sigma\rho}_{\beta\delta}U^\delta_{\gamma\alpha}\bigg]\varepsilon^{ab}
 \hat{T}^{\alpha\beta}\pi^\gamma\\&& 
  -\lambda_\rho\bigg[\{\{V^\rho_\gamma,T_\beta\},T_\alpha\}-\frac{1}{2}\bigg(
\{V^\rho_\gamma,T_\delta\}+(-1)^{\varepsilon_\gamma\varepsilon_\delta}\{V^\rho_
 \delta,T_\gamma\}\bigg)U^\delta_{\beta\alpha}\\&& 
  -\frac{1}{2}U^\rho_{\alpha\delta}\{V^\delta_\gamma,T_\beta\}(-1)^
 {\varepsilon_\alpha(\varepsilon_\beta+\varepsilon_\gamma)}-\frac{1}{3}\{H_0,
\{U^\rho_{\beta\alpha},T_\gamma\}\}(-1)^{\varepsilon_\gamma(\varepsilon_\alpha+
 \varepsilon_\beta)}\\&& 
  -\frac{1}{6}\{H_0,U^\rho_{\gamma_\delta}U^\delta_{\beta\alpha},\}-
 \frac{1}{3}V^\rho_\delta\bigg(\{U^\delta_{\beta\alpha},T_\gamma\}(-1)^
 {\varepsilon_\gamma(\varepsilon_\alpha+\varepsilon_\beta)}+\frac{1}{2}U^
 \delta_{\gamma\sigma}U^\sigma_{\beta\alpha}\bigg)\\&& 
  +\frac{1}{3}\bigg(\{U^\rho_{\beta\delta},T_\gamma\}
 (-1)^{\varepsilon_\gamma(\varepsilon_\beta+\varepsilon_\delta)}+\frac{1}{2}U^
 \rho_{\gamma\sigma}U^\sigma_{\beta\delta}\bigg)V^\delta_\alpha\\&& 
  +\frac{1}{3}\bigg(\{U^\rho_{\gamma\alpha},T_\delta\}(-1)^{\varepsilon_\alpha
 \varepsilon_\delta}+(-1)^{\varepsilon_\gamma\varepsilon_\delta}\{U^\rho_
 {\delta\alpha},T_\gamma\}(-1)^{\varepsilon_\alpha\varepsilon_\gamma}\\&& 
  +\frac{1}{2}U^\rho_{\gamma\sigma}U^\sigma_{\delta\alpha}+\frac{1}{2}(-1)^
 {\varepsilon_\gamma\varepsilon_\delta}U^\rho_{\delta\sigma}U^\sigma_{\gamma
 \alpha}\bigg)V^\delta_\beta(-1)^{\varepsilon_\alpha(\varepsilon_\beta+
 \varepsilon_\delta)}\bigg]C^{\alpha a}\hat{T}^{\beta\gamma}\\&& 
  -\frac{1}{2}\lambda_\rho\lambda_\sigma\bigg[\{V^\sigma_\gamma,U^\rho_
 {\beta\alpha}\}+U^\sigma_{\gamma\delta}E^{\delta\rho}_{\beta\alpha}\bigg](-1)^
 {\varepsilon_\beta+\varepsilon_\gamma(\varepsilon_\rho+1)}C^{\alpha a}\pi^
 \beta\pi^\gamma. 
\end{eqnarray*}
 Eqs. (2.11) for ${\cal H}_3$ are solved by the method given above. It implies
 consideration of summands present in $D^a_3$, taking into account the gauge
 algebra structural relations (2.33), (2.34), (2.40), (2.42), and the
properties
 of structural coeffitients (2.2), (2.32), (2.41) with allowance made for the
 following identity
\[
 \varepsilon_{ab}\bigg(C^{\alpha a}C^{\beta b}C^{\gamma c}+{\rm
cycl.perm.}(abc)
 \bigg)\equiv 0.
\]
 Omitting details of calculation, we give here the resultant form of
 ${\cal H}_3$
\begin{eqnarray}
 {\cal H}_3&=&\frac{1}{3}{\cal P}_{\delta a}{\cal P}_{\rho b}{\cal P}_
 {\sigma c}(-1)^{\varepsilon_\rho}E^{\sigma\rho\delta}_{\gamma\beta\alpha}T^
 {\alpha\beta\gamma\{abc\}}\nonumber\\&& 
  +\lambda_\delta{\cal P}_{\rho a}{\cal P}_{\sigma b}
 (-1)^{\varepsilon_\rho}E^{\sigma\rho\delta}_{\gamma\beta\alpha}(-1)^
 {\varepsilon_\alpha\varepsilon_\gamma}T^{\alpha\beta\{ab\}}\pi^\gamma\nonumber
 \\&& 
  +\frac{1}{3}\lambda_\rho{\cal P}_{\sigma a}(-1)^{\varepsilon_\rho}\bigg[E^
 {\sigma\rho}_{\gamma\delta}U^\delta_{\beta\alpha}+2\bigg(\{E^{\sigma\rho}_
{\gamma\alpha},T_\beta\}(-1)^{\varepsilon_\alpha\varepsilon_\beta}\nonumber\\&&
  -U^\sigma_{\gamma\delta}E^{\delta\rho}_{\beta\alpha}(-1)^
 {\varepsilon_\gamma\varepsilon_\rho}\bigg)-U^\rho_{\gamma\delta}E^
 {\delta\sigma}_{\beta\alpha}(-1)^{\varepsilon_\sigma(\varepsilon_\gamma+
 \varepsilon_\rho)}\nonumber\\&& 
 -\{V^\sigma_\gamma,U^\rho_{\beta\alpha}\}(-1)^{\varepsilon_\gamma\varepsilon_
 \rho}-2\{V^\rho_\gamma,U^\sigma_{\beta\alpha}\}(-1)^{\varepsilon_\sigma(
\varepsilon_\gamma+\varepsilon_\rho)}\bigg](-1)^{\varepsilon_\beta+\varepsilon_\gamma}C^{\alpha a}\hat{T}^{\beta\gamma}
 \nonumber\\&& 
 +\lambda_\rho\lambda_\sigma\bigg[\{V^\sigma_\beta,U^\rho_{\gamma\alpha}\}+U^
 \sigma_{\beta\delta}E^{\delta\rho}_{\gamma\alpha}\bigg](-1)^{\varepsilon_
 \beta(\varepsilon_\gamma+\varepsilon_\rho)}\hat{T}^{\alpha\beta}\pi^\gamma. 
\end{eqnarray}
 In view of (2.43) and the results obtained above (2.23), (2.25), (2.38),
 (2.39), we conclude that in the framework of the Sp(2)-covariant version of
 generalized canonical quantization, as well as in its standard formulation,
 solutions to the generating equations are completely defined by the gauge
 algebra structural coeffitients.
 The boson function
\begin{eqnarray*}
 {\cal H}(P_A,Q^A)&=&H_0(p_\imath,q^\imath)+{\cal H}_1(P_A,Q^A)+
 {\cal H}_2(P_A,Q^A)\\&&
 +{\cal  H}_3(P_A,Q^A)+O(C^{4-n}\pi^n),\;\;\;\;0\leq n\leq4
\end{eqnarray*}
 and the fermion functions
\begin{eqnarray*}
\Omega^a(P_A,Q^A)&=&\Omega^a_1(P_A,Q^A)+\Omega^a_2(P_A,Q^A)+\Omega^a_3(P_A,Q^A)
 \\&&+O(C^{4-n}\pi^n),\;\;\;\;0\leq n\leq4
\end{eqnarray*}
 satisfy the gauge algebra generating equations (2.4) up to the 3d order
 inclusively. As regards the description of arbitrariness in the solutions to
 the generating equations, note that this question has been thoroughly studied
 in Ref. 18.

 Concluding, let us consider the special case of dynamical systems with
 constant structural coeffitients, such that
\[
 \{V^\gamma_\alpha,T_\beta\}=\{U^\delta_{\alpha\beta},T_\gamma\}=\{U^\delta_
 {\alpha\beta},V^\rho_\gamma\}=0,
\]
 given this, we assume that
\[
 E^{\gamma\delta}_{\alpha\beta}=E^{\delta\rho}_{\alpha\beta\gamma}=E^{\delta
 \rho\sigma}_{\alpha\beta\gamma}=0.
\]
 Then the functions $\cal H$ and $\Omega^a$ take on the form
\begin{eqnarray*}
 {\cal H}={\cal H}_{[1]}=H_0+{\cal P}_{\beta a}V^\beta_\alpha C^{\alpha a}+
 \lambda_\beta V^
 \beta_\alpha\pi^\alpha,
\end{eqnarray*}
\begin{flushright}(2.44)\end{flushright}
\begin{eqnarray*}
 \Omega^a\;=\;\Omega^a_{[3]}&=&T_\alpha C^{\alpha a}+\varepsilon^{ab}{\cal
P}_{\alpha b}\pi^\alpha+
 \frac{1}{2}{\cal P}_{\gamma b}U^\gamma_{\beta\alpha}T^{\alpha\beta\{ab\}}-
 \frac{1}{2}\lambda_\gamma U^\gamma_{\beta\alpha}(-1)^{\varepsilon_\beta}C^
 {\alpha a}\pi^\beta\\&&
 +\frac{1}{6}\lambda_\rho U^\rho_{\gamma\delta}U^\delta_{\beta\alpha}(-1)^
 {\varepsilon_\beta+\varepsilon_\gamma}C^{\alpha a}\hat{T}^{\beta\gamma}.
\end{eqnarray*}
 In a particular case of dynamical systems of rank 1 with a closed algebra
 where all the constraints $T_\alpha$ are boson functions (i.~e. $\varepsilon_
 \alpha=0$), this result coincide with the exact solution to the generating
 equations for the functions $\cal H$ and $\Omega^a$, obtained in Ref.~12.\\
[1mm]
\setcounter{sct}{3}
\setcounter{equation}{0}
{\bf 3. The Solution of Equations for S}\\[0.1mm]

Now, turning to the Lagrangian version of the Sp(2)-covariant quantization
method, consider the classical theory of fields $A^\imath$,
$\varepsilon(A^\imath)\equiv \varepsilon_i$, described by the action
${\cal L}={\cal L}(A)$ invariant under the gauge transformations
$\delta A^\imath ={\cal R}^{\imath}_{\alpha} (A)\xi^{\alpha}$
\begin{eqnarray} {\cal L},_i(A){\cal R}^\imath_{\alpha}(A) = 0\; ,
\end{eqnarray}
where ${\cal R}^{\imath}_{\alpha}(A)$ are generators of gauge transformations
$\varepsilon ({\cal R}^{\imath}_{\alpha}(A)) = \varepsilon_\imath+\varepsilon_
\alpha$ and $\xi^\alpha$ are arbitrary functions $\varepsilon(\xi^\alpha)
\equiv\varepsilon_\alpha$.
Let us now introduce, in accordance with Refs.15, 17, the total configuration
space $\Phi^A \; (\varepsilon(\Phi^A)\equiv\varepsilon^A)$ whose manifest
structure is determined by the fact whether the set of generators is lineary
independent (irreducible theories) or lineary dependent (reducible theories).
In what follows we shall restrict ourselves to consideration of irreducible
theories.
The total configuration space of the theories in question has the following
structure
\begin{eqnarray}
& &\qquad \Phi^A = (A^{\imath}, B^{\alpha}, C^{\alpha a}) \;, \nonumber
\\
& &{\rm ngh}(A^\imath) = 0\;,\; {\rm ngh}(C^{\alpha a}) = 1\;,\; {\rm
ngh}(B^{\alpha}) = 2
\end{eqnarray}
on account of extention of the initial configuration space $A^\imath$ by
introducing additional fields $B^{\alpha}\; (\varepsilon(B^\alpha)=\varepsilon_
\alpha)$
and Sp(2) doublets of ghost fields $C^{\alpha a}\; (\varepsilon(C^{\alpha a})=
\varepsilon_\alpha+1)$.
We also introduce the sets of anifields $\Phi^{*}_{Aa}$ and $\bar {\Phi}_A$
\begin{eqnarray*}
& &\Phi^{*}_{Aa} = (A^{*}_{\imath a},\; B^{*}_{\alpha a},\; C^{*}_{\alpha a
b})\;,\; \bar {\Phi}_A = (\bar {A}_{\imath},\; \bar {B}_{\alpha},\; \bar
{C}_{\alpha a}) \;,
\\
& &\qquad \varepsilon (\Phi^{*}_{Aa}) = \varepsilon_A + 1\;,\; \varepsilon(\bar
{\Phi}_A) = \varepsilon_A \;,
\\
& &{\rm ngh}(\Phi^{*}_{Aa}) = -1-{\rm ngh}(\Phi^A)\;,\; {\rm ngh}(\bar
{\Phi}_A) = -2-{\rm ngh}(\Phi^A)\;.
\end{eqnarray*}
The antifields $\Phi^{*}_{Aa}$ play the role of sources of BRST and anti-BRST
transformations, whereas the antifields $\bar {\Phi}_A$ are the sources of
mixed BRST and anti-BRST transformations.

The basic object in the Lagrangian version of Sp(2)-covariant
${\rm quanti-}$\\${\rm zation}^{15,17}$ is the boson functional $S$, which
satisfies the following generating
equations
\begin{eqnarray}
 \frac {1}{2}(S,S)^a + V^aS = i\hbar\Delta^\alpha S
\end{eqnarray}
with the boundary condition
\begin{eqnarray}
 S_{\mid {\Phi^{*}_{a} = \bar {\Phi} = \hbar = 0}} ={\cal L}(A)\;.
\end{eqnarray}
In (3.3) we used designation $(\; ,\; )^a$ for an extended ${\rm
antibracket}^{15}$ introduced
for two arbitrary functionals $F$ and $G$ by the rule
\begin{eqnarray}
(F,G)^a = \frac {\delta F}{\delta \Phi^A} \frac {\delta G}{\delta \Phi^{*}_
{Aa}} -
\frac {\delta G}{\delta \Phi^A} \frac {\delta F}{\delta \Phi^{*}_{Aa}}(-1)^
{(\varepsilon(F)+1)(\varepsilon(G)+1)}\; ,
\end{eqnarray}
$V^a$ and $\Delta^a$ are operators defined in the form
\begin{eqnarray*}
 V^a = \varepsilon^{ab}\Phi^{*}_{Ab} \frac {\delta  }{\delta \bar\Phi_A}\;,\;\;
 \Delta^a = (-1)^{\varepsilon_A}\frac {\delta_l}{\delta \Phi^A} \frac {\delta
}{\delta \Phi^{*}_{Aa}}\;,
\end{eqnarray*}
$\hbar$ is the Plank constant. The algebra of operators $V^a$ and $\Delta^a$
as well as the properties of the extended antibracket are studied in
detail in Ref.~15, and we shall not discuss here these questions.

The generating functional of the Green's functions $Z(J)$ for the fields of the
extended configuration space is representable in the ${\rm form}^{15}$
\begin{eqnarray}
 Z(J) &=& \int d\Phi\; d\Phi^{*}_a\;d\bar{\Phi}\;d\lambda\;d\Pi^a\;\exp\bigg
 \lbrace\frac {i}{\hbar}\bigg(S(\Phi,\Phi^{*}_a,\bar\Phi)+\Phi^{*}_{Aa}
 \Pi^{Aa}\nonumber\\&&
 + \bigg(\bar\Phi_A - \frac {\delta F}{\delta \Phi^A}\bigg)\lambda^A
 - \frac {1}{2} \varepsilon_{ab}\Pi^{Aa}{\frac {\delta^2 F}{\delta \Phi^A
\delta
 \Phi^B}}\Pi^{Bb}+J_A\Phi^A\bigg)\bigg\rbrace\; ,
\end{eqnarray}
where $J_A$ are the usual sources to the fields $\Phi^A$
$(\varepsilon(J_A)=\varepsilon_A)$,
$\Pi^{Aa}$ and $\lambda^A$ are the sets of auxiliary fields
$(\varepsilon(\Pi^{Aa})=\varepsilon_A+1 \;,\;
\varepsilon(\lambda^A)=\varepsilon_A)$, $F=F(\Phi)$ is the boson gauge
functional,
$S(\Phi,\Phi^{*}_a\;,\;\bar\Phi)$ is a solution to Eqs.(3.3) with the boundary
condition
(3.4).

An impotant property of the integrand in Eqs.(3.6) for $J_A=0$ is its
invariance
under the following transformations of global supersymmetry
\begin{eqnarray}
 \delta\Phi^A &=& \Pi^{Aa}\mu_a\;,\; \delta\Phi^{*}_{Aa}=\mu_a\frac {\delta
S}{\delta\Phi^A}\;,\;
 \delta\bar\Phi_A=\varepsilon^{ab}\mu_a\Phi^{*}_{Ab}\;,  \nonumber
\\
& &\delta\Pi^{Aa}=-\varepsilon^{ab}\lambda^A\mu_b,\; \delta\lambda^A=0\;,
\end{eqnarray}
where $\mu_a$ is a doublet of constant anticommuting Grassmann parameters.
The transformations (3.7) realize the extended BRST transformations in the
space of the variables
$\Phi,\; \Phi^{*}_a,\; \bar\Phi,\; \Pi^a,\; \lambda$.

The symmetry of the vacuum functional $Z(0)$ under the transformations (3.7)
permits establishing the independence of the S matrix on the choice of a gauge.
Indeed, suppose $Z_F \equiv Z(0)$. We shall now change the gauge $F\to F+
\Delta F$.
In the functional integral for $Z_{F+\Delta F}$ we make the change of variables
(3.7), choosing for the parameters $\mu_a$
\begin{eqnarray*}
 \mu_a = \frac {i}{2\hbar}\varepsilon_{ab} \frac {\delta(\Delta F)}{\delta
\Phi^A}\Pi^{Ab}\;,
\end{eqnarray*}
we find that $Z_{F+\Delta F}=Z_F$ and therefore the S matrix is gauge
invariant.
Next, the extended BRST symmetry permits deriving the Ward identities for the
generating functional of the vertex functions (the effective action) $\Gamma=
\Gamma(\Phi,\Phi^{*}_a,\bar\Phi)$
\begin{eqnarray*}
 \frac {1}{2}(\Gamma,\Gamma)^a + V^a\Gamma = 0\;.
\end{eqnarray*}
The identities for $\Gamma$ provide a basis for establishing the gauge
invariant
renormalizability of the general gauge theories in the framework of the Sp(2)-
covariant ${\rm quantization.}^{17}$

In Refs. 15, 17 there is proved the existence theorem for solution to Eqs.
(3.3)
in the form of expansion in power series of $\hbar$
\begin{eqnarray*}
 S(\Phi,\Phi^{*}_a,\bar\Phi)=\sum_{n=0}^\infty \hbar^n
S^{(n)}(\Phi,\Phi^{*}_a,\bar\Phi)\;,
\end{eqnarray*}
where the functionals $S^{(n)}(\Phi,\Phi^{*}_a,\bar\Phi)$ are supposed to be
regular with respect to all the variables. In particular, the tree (classical)
approximation $S^{(0)}$ satisfies the following generating equations
\begin{eqnarray}
 \frac {1}{2}(S^{(0)},S^{(0)})^a + V^aS^{(0)} = 0\;.
\end{eqnarray}
In what follows we shall concentrate our attention upon Eqs. (3.8). One can
seek solution
to Eqs. (3.8) in the form of expansion in power series of ghost and auxiliary
fields $C^{\alpha a},\; B^\alpha$
\begin{eqnarray}
 S^{(0)}(\Phi,\Phi^{*}_a,\bar\Phi) = {\cal {L}}(A) + \sum_{n=1}^\infty S_n
(\Phi,\Phi^{*}_a,\bar\Phi)\;,
\end{eqnarray}
having required the new ghost number and the Grassmann parity to be conserved
in every order of perturbation series
\begin{eqnarray*}
 \varepsilon (S_n) = {\rm ngh}(S_n) = 0\;.
\end{eqnarray*}
In (3.9) $S_n (\Phi,\Phi^{*}_a,\bar\Phi)$ are some $n$th order polinomials in
the
fields $C^{\alpha a},\; B^\alpha$. The requirement of the new ghost number
conservation
leads to the fact that $S_n$ must be polinomials in the antifields
$\Phi^{*}_{Aa}, \bar\Phi_A$.  In Ref. 15 it is shown that the
first approximation $S_1$  could be chosen in the form
\begin{eqnarray} S_1 (\Phi,\Phi^{*}_a,\bar\Phi) = A^{*}_{\imath a}
 {\cal R}^{\imath}_{\alpha}(A) C^{\alpha a} + \bar {A}_\imath {\cal
 R}^{\imath}_{\alpha}(A) B^\alpha - \varepsilon^{ab} C^{*}_{\alpha ab}
 B^\alpha \;.  \end{eqnarray} Then the higher approximations in (3.9)
are sought from the equations \begin{eqnarray} W^a S_{n+1} =
 F^a_{n+1}\;,\; n\geq 1\;, \end{eqnarray} where the operators $W^a$ are
defined in the form \begin{eqnarray} W^a &=& {\cal L},_\imath \frac
 {\delta}{\delta A^{*}_{\imath a}} + A^{*}_{\imath b} {\cal
 R}^{\imath}_{\alpha} \frac {\delta}{\delta C^{*}_{\alpha ab}} +
 \bigg(\bar {A}_\imath {\cal R}^{\imath}_{\alpha} - \varepsilon^{bc}
C^{*}_{\alpha bc}\bigg) \frac {\delta}{\delta B^{*}_{\alpha a}} \nonumber
\\
& & + (-1)^{\varepsilon_\alpha}\varepsilon^{ab} B^\alpha \frac
{\delta_l}{\delta C_{\alpha b}} +
 V^a
\end{eqnarray}

and possess the properties
\begin{eqnarray*}
 W^{{\{a}} W^{b\}} = 0\;.
\end{eqnarray*}
The functionals $F^a_{n+1}$ are constucted from $S_k,\; k\leq n$ by the rule
\begin{eqnarray}
 F^a_{n+1} = -\frac{1}{2}(S_{[n]},S_{[n]})^a_{n+1} \;,
\end{eqnarray}
where
\begin{eqnarray*}
 S_{[n]} = {\cal L}(A) + \sum_{k=1}^n S_k\;,
\end{eqnarray*}
the symbol $(\; ,\; )^a_k$ denotes the $k$th order for the extended antibracket
in power
series of the fields $C^{\alpha a}$ and $B^\alpha$. The functionals $F^a_{n+1}$
satisfy the equations
\begin{eqnarray*}
 W^{\{a}_{\;} F^{b\}}_{n+1} = 0\;,
\end{eqnarray*}
being the compability conditions for Eqs. (3.11).

Let us obtain the explicit solutions to Eqs. (3.11) when $n=1,2$ in the case
of general gauge theories described by the action $\cal {L}(A)$ invariant
under the gauge transformations (3.1). The algebra of generators ${\cal
R}^{\imath}_{\alpha}(A)$
has the following general form
\begin{eqnarray}
& &{\cal R}^{\imath}_{\alpha ,\jmath}(A){\cal R}^{\jmath}_{\beta}(A) -
 (-1)^{\varepsilon_\alpha\varepsilon_\beta}{\cal R}^{\imath}_{\beta
,\jmath}(A){\cal R}^{\jmath}_{\alpha}(A)=  \nonumber
\\
& &\quad -{\cal R}^{\imath}_{\gamma}(A){\cal F}^{\gamma}_{\alpha\beta}(A) -
 {\cal L},_{\jmath}(A) M^{\imath\jmath}_{\alpha\beta}(A)\;,
\end{eqnarray}
where the structural coefficients ${\cal F}^{\gamma}_{\alpha\beta}(A)$ and
$M^{\imath\jmath}_{\alpha\beta}(A)$ possess the properties of generalized
antisymmetry
\begin{eqnarray}
& &\qquad {\cal F}^{\gamma}_{\alpha\beta}(A) =
-(-1)^{\varepsilon_\alpha\varepsilon_\beta}
 {\cal F}^{\gamma}_{\beta\alpha}(A)\;,\nonumber
\\
& & M^{\imath\jmath}_{\alpha\beta}(A) =
 -(-1)^{\varepsilon_\alpha\varepsilon_\beta}M^{\imath\jmath}_{\beta\alpha}(A) =
-(-1)^{\varepsilon_\imath\varepsilon_\jmath}M^{\jmath\imath}_{\alpha\beta}(A)\;.
\end{eqnarray}

Consider here the solution of Eqs. (3.11) for $S_2$. To do this we shall make
use of the decomposition (2.18) of an Sp(2) tensor of rank two, constructed
from $C^{\alpha a}$. Then, by virtue of the properties (2.19) and the algebra
of generators (3.14), we find the functionals $F^a_2$ in the form
\begin{eqnarray}
& & F^a_2 = -\frac{1}{2}A^{*}_{\imath d}\bigg({\cal
R}^{\imath}_{\gamma}(A){\cal F}^{\gamma}_{\alpha\beta}(A) +
 {\cal L},_\jmath (A)
M^{\imath\jmath}_{\alpha\beta}(A)\bigg)T^{\beta\alpha\{ad\}}   \nonumber
\\
& &\qquad + \frac{1}{2}\bar {A}_{\imath }\bigg({\cal
R}^{\imath}_{\gamma}(A){\cal F}^{\gamma}_{\alpha\beta}(A) +
 {\cal L},_\jmath (A) M^{\imath\jmath}_{\alpha\beta}(A)\bigg)C^{\beta
a}B^\alpha(-1)^{\varepsilon_\alpha}   \nonumber
\\
& &\qquad + \frac {1}{2}\varepsilon^{ad}A^{*}_{\imath
d}N^\imath_{\alpha\beta}(A)\hat{T}^{\beta\alpha} -
  \frac {1}{2}\bar{A}_{\imath}N^\imath_{\alpha\beta}(A)C^{\beta
a}B^\alpha(-1)^{\varepsilon_\alpha} \;,
\end{eqnarray}
where
\begin{eqnarray*}
& & N^\imath_{\alpha\beta}(A) = {\cal R}^{\imath}_{\alpha ,\jmath}(A){\cal
R}^{\jmath}_{\beta}(A) +
 (-1)^{\varepsilon_\alpha\varepsilon_\beta}{\cal R}^{\imath}_{\beta
,\jmath}(A){\cal R}^{\jmath}_{\alpha}(A)\;,
\\
& &\qquad N^\imath_{\alpha\beta}(A) =
(-1)^{\varepsilon_\alpha\varepsilon_\beta}N^\imath_{\beta\alpha}(A)\;.
\end{eqnarray*}
Solving Eqs. (3.11) for $S_2$ is as follows. Consider the operators $W^a$ to
act
upon the functional
\[
 \frac{1}{2}C^{*}_{\gamma bd}{\cal
F}^{\gamma}_{\alpha\beta}(A)T^{\beta\alpha\{bd\}}
\]
\begin{eqnarray}
& & W^a\bigg(-\frac{1}{2}C^{*}_{\gamma bd}{\cal
F}^{\gamma}_{\alpha\beta}(A)T^{\beta\alpha\{bd\}}\bigg) =
 -\frac{1}{2}A^{*}_{\imath d}{\cal R}^{\imath}_{\gamma}(A){\cal
F}^{\gamma}_{\alpha\beta}(A)
 T^{\beta\alpha\{ad\}}   \nonumber
\\
& &\qquad\qquad - \frac{1}{2}\varepsilon^{ab}\bigg(C^{*}_{\gamma bd} +
C^{*}_{\gamma db}\bigg){\cal F}^{\gamma}_{\alpha\beta}(A)B^{\beta}C^{\alpha
d}\;.
\end{eqnarray}
Hence we find that the first summand in (3.17) present in (3.16) as well.
In a similar way, the structure of the form
\[
-\frac{1}{2}A^{*}_{\imath d}{\cal L},_\jmath
(A)M^{\imath\jmath}_{\alpha\beta}(A)T^{\beta\alpha\{ad\}}
\]
is reprodused by applying the operators $W^a$ to the functional
\[
\frac{1}{4}A^{*}_{\imath b}A^{*}_{\jmath
d}(-1)^{\varepsilon_\imath}M^{\imath\jmath}_{\alpha\beta}(A)T^{\beta\alpha\{bd\}}\;,
\]
namely
\begin{eqnarray}
 & &W^a\bigg(\frac{1}{4}A^{*}_{\imath b}A^{*}_{\jmath
d}(-1)^{\varepsilon_\imath}M^{\imath\jmath}_{\alpha\beta}(A)T^{\beta\alpha\{bd\}}\bigg) =
 -\frac{1}{2}A^{*}_{\imath d}{\cal L},_\jmath
M^{\imath\jmath}_{\alpha\beta}(A)T^{\beta\alpha\{ad\}}    \nonumber
\\
 & &\qquad\qquad+\frac{1}{2}\varepsilon^{ab}A^{*}_{\imath b}A^{*}_{\jmath
d}(-1)^{\varepsilon_\imath}M^{\imath\jmath}_{\alpha\beta}(A)B^{\beta}C^{\alpha
d}\;.
\end{eqnarray}
Treatment of the terms
\[
\frac{1}{2}\bar {A}_{\imath }{\cal R}^{\imath}_{\gamma}(A){\cal
F}^{\gamma}_{\alpha\beta}(A)C^{\beta a}B^\alpha(-1)^{\varepsilon_\alpha}
\]
and
\[
\frac{1}{2}\bar {A}_{\imath}{\cal L},_\jmath
M^{\imath\jmath}_{\alpha\beta}(A)C^{\beta a}B^\alpha(-1)^{\varepsilon_\alpha}
\]
present in $F^a_2$ consists in consideration of $W^a$ to act respectively
upon
\[
\frac{1}{2}B^{*}_{\gamma b}{\cal F}^{\gamma}_{\alpha\beta}(A)C^{\beta
b}B^\alpha(-1)^{\varepsilon_\alpha}
\]
and
\[
\frac{1}{2}\bar {A}_{\imath}A^{*}_{\jmath
b}(-1)^{\varepsilon_\imath}M^{\imath\jmath}_{\alpha\beta}(A)C^{\beta
b}B^\alpha(-1)^{\varepsilon_\alpha}\;,
\]
namely
\begin{eqnarray}
 & &W^a\bigg(\frac{1}{2}B^{*}_{\gamma b}{\cal
F}^{\gamma}_{\alpha\beta}(A)C^{\beta b}B^\alpha(-1)^{\varepsilon_\alpha}\bigg)
=
 \frac{1}{2}\bar {A}_{\imath }{\cal R}^{\imath}_{\gamma}(A){\cal
F}^{\gamma}_{\alpha\beta}(A)C^{\beta a}B^\alpha(-1)^{\varepsilon_\alpha}
\nonumber
\\
 & &\qquad\qquad - \frac{1}{2}\varepsilon^{bd}C^{*}_{\gamma bd}{\cal
F}^{\gamma}_{\alpha\beta}(A)C^{\beta a}B^{\alpha}(-1)^{\varepsilon_\alpha}\;,
\end{eqnarray}
\begin{eqnarray}
 & &W^a\bigg(\frac{1}{2}\bar {A}_{\imath}A^{*}_{\jmath
b}(-1)^{\varepsilon_\imath}M^{\imath\jmath}_{\alpha\beta}(A)C^{\beta
b}B^\alpha(-1)^{\varepsilon_\alpha}\bigg) = \nonumber \\ &&
 \qquad\qquad\frac{1}{2}\bar {A}_{\imath}{\cal L},_\jmath (A)
M^{\imath\jmath}_{\alpha\beta}(A)C^{\beta a}B^\alpha(-1)^{\varepsilon_\alpha}
\nonumber
\\
 & &\qquad\qquad- \frac{1}{2}\varepsilon^{ab}A^{*}_{\imath b}A^{*}_{\jmath
d}(-1)^{\varepsilon_\imath}M^{\imath\jmath}_{\alpha\beta}(A)B^{\beta}C^{\alpha
d} \;.
\end{eqnarray}
Consider the functional of the form
\begin{eqnarray}
 X_1(\Phi^A,\Phi^{*}_{Aa},\bar\Phi_A) &=& -\frac{1}{2}C^{*}_{\gamma bd}{\cal
F}^{\gamma}_{\alpha\beta}(A)T^{\beta\alpha\{bd\}} +
 \frac{1}{2}B^{*}_{\gamma b}{\cal F}^{\gamma}_{\alpha\beta}(A)C^{\beta
b}B^\alpha(-1)^{\varepsilon_\alpha}  \nonumber
\\
 & & +\frac{1}{4}A^{*}_{\imath b}A^{*}_{\jmath
d}(-1)^{\varepsilon_\imath}M^{\imath\jmath}_{\alpha\beta}(A)T^{\beta\alpha\{bd\}}  \nonumber
\\
 & & +\frac{1}{2}\bar {A}_{\imath}A^{*}_{\jmath
b}(-1)^{\varepsilon_\imath}M^{\imath\jmath}_{\alpha\beta}(A)C^{\beta
b}B^\alpha(-1)^{\varepsilon_\alpha}\;,
\end{eqnarray}
then
\begin{eqnarray}
 & &W^a X_1 = -\frac{1}{2}A^{*}_{\imath d}{\cal L},_\jmath
M^{\imath\jmath}_{\alpha\beta}(A)T^{\beta\alpha\{ad\}} +
 \frac{1}{2}\bar {A}_{\imath}{\cal L},_\jmath (A)
M^{\imath\jmath}_{\alpha\beta}(A)C^{\beta a}B^\alpha(-1)^{\varepsilon_\alpha}
\nonumber
\\
 & &\qquad-\frac{1}{2}A^{*}_{\imath d}{\cal R}^{\imath}_{\gamma}(A){\cal
F}^{\gamma}_{\alpha\beta}(A)T^{\beta\alpha\{ad\}} +
 \frac{1}{2}\bar {A}_{\imath }{\cal R}^{\imath}_{\gamma}(A){\cal
F}^{\gamma}_{\alpha\beta}(A)C^{\beta a}B^\alpha(-1)^{\varepsilon_\alpha}
\nonumber
\\
 & &\qquad\qquad\qquad- \frac{1}{2}\varepsilon^{ab}(C^{*}_{\gamma bd} +
C^{*}_{\gamma db}){\cal F}^{\gamma}_{\alpha\beta}(A)B^{\beta}C^{\alpha d}
\nonumber
\\
 & &\qquad\qquad\qquad - \frac{1}{2}\varepsilon^{bd}C^{*}_{\gamma bd}{\cal
F}^{\gamma}_{\alpha\beta}(A)C^{\beta a}B^{\alpha}(-1)^{\varepsilon_\alpha}\;.
\end{eqnarray}
The functional $X_1(\Phi^A,\Phi^{*}_{Aa},\bar\Phi_A)$ possesses the properties
$\varepsilon(X_{1})={\rm ngh}(X_{1})=0$ with the multipliers chosen in such a
way
that $W^aX_1$ should exhaust the summands in $F^a_2$ described above. In
calculating
$W^aX_1$ (3.22) one has to make use of the generalized antisymmetry properties
(3.15) of the gauge algebra structural coeffitients ${\cal
F}^{\gamma}_{\alpha\beta}(A)$
and $M^{\imath\jmath}_{\alpha\beta}(A)$ with allowance made for the
properties (2.19) and for the definition of the tensors
$T^{\beta\alpha\{ad\}}$, $\varepsilon^{ab}\hat{T}^{\beta\alpha}$
(2.18) and the properties (2.19). In particular,
\begin{eqnarray}
 {\cal F}^{\gamma}_{\alpha\beta}(A)B^{\alpha}B^{\beta}\equiv
M^{\imath\jmath}_{\alpha\beta}(A)B^{\alpha}B^{\beta}\equiv 0\;.
\end{eqnarray}
Comparison of $W^a X_1$ with $F^a_2$ implies
\begin{eqnarray}
 F^a_2 - W^a X_1 &=&
  \frac {1}{2}\varepsilon^{ad}A^{*}_{\imath
d}N^\imath_{\alpha\beta}(A)\hat{T}^{\beta\alpha} -
  \frac {1}{2}\bar{A}_{\imath}N^\imath_{\alpha\beta}(A)C^{\beta
a}B^\alpha(-1)^{\varepsilon_\alpha}  \nonumber
\\
 & &+\varepsilon^{ab}C^{*}_{\gamma bd} {\cal
F}^{\gamma}_{\alpha\beta}(A)B^{\beta}C^{\alpha d}\;.
\end{eqnarray}
In establishing (3.24) we made use of the identity
\begin{eqnarray*}
 & & \frac{1}{2}\varepsilon^{ab}\bigg(C^{*}_{\gamma bd} + C^{*}_{\gamma
db}\bigg){\cal F}^{\gamma}_{\alpha\beta}(A)B^{\beta}C^{\alpha d}
 + \frac{1}{2}\varepsilon^{bd}C^{*}_{\gamma bd}{\cal
F}^{\gamma}_{\alpha\beta}(A)B^{\beta }C^{\alpha a}=
\\
 & &\qquad\qquad\qquad \varepsilon^{ab}C^{*}_{\gamma bd} {\cal
F}^{\gamma}_{\alpha\beta}(A)B^{\beta}C^{\alpha d}\;,
\end{eqnarray*}
verified directly.
Making an analysis similar to the one discussed above for each summand present
in (3.24), consider $W^a$ to act upon the functional $X_2$ of the form
\begin{eqnarray}
 X_2 = \frac
{1}{2}\bar{A}_{\imath}N^\imath_{\alpha\beta}(A)\hat{T}^{\beta\alpha} +
 \bar{C_{\gamma d}}{\cal F}^{\gamma}_{\alpha\beta}(A)B^{\beta}C^{\alpha d}\;.
\end{eqnarray}
One readily finds that $F^a_2 - W^a(X_1 + X_2) = 0$. Consequently, the solution
to Eqs. (3.11) with $F^a_2$ (3.16) has the form
\begin{eqnarray}
 S_2 &=& X_1 + X_2 = -\frac{1}{2}C^{*}_{\gamma bd}{\cal
F}^{\gamma}_{\alpha\beta}(A)T^{\beta\alpha\{bd\}} +
 \frac{1}{2}B^{*}_{\gamma b}{\cal F}^{\gamma}_{\alpha\beta}(A)C^{\beta
b}B^\alpha  \nonumber
\\
 & &\qquad\qquad +\frac{1}{4}A^{*}_{\imath b}A^{*}_{\jmath
d}(-1)^{\varepsilon_\imath}M^{\imath\jmath}_{\alpha\beta}(A)T^{\beta\alpha\{bd\}}  \nonumber
\\
 & &\qquad\qquad+ \frac{1}{2}\bar {A}_{\imath}A^{*}_{\jmath
b}(-1)^{\varepsilon_\imath}M^{\imath\jmath}_{\alpha\beta}(A)C^{\beta
b}B^\alpha(-1)^{\varepsilon_\alpha}  \nonumber
\\
 & &\qquad\qquad+ \frac
{1}{2}\bar{A}_{\imath}N^\imath_{\alpha\beta}(A)\hat{T}^{\beta\alpha} +
 \bar{C_{\gamma d}}{\cal F}^{\gamma}_{\alpha\beta}(A)B^{\beta}C^{\alpha d}\;.
\end{eqnarray}

Consider now the solution of Eqs. (3.11) for $S_3$. It is convinient, as
mentioned above, to seek the solution with allowance made for the decomposition
(2.27) of an Sp(2) tensor of rank three constructed from $C^{\alpha a}$.

In solving Eqs. (3.11) for $S_3$ it is necessary to employ subsequent
structural relations of the gauge ${\rm algebra}^{23}$, that is to say,
the generalized Jacobi identity
\begin{eqnarray}
 {\cal R}^{\imath}_{\gamma}(A)D^{\gamma}_{\alpha\beta\delta}(A) +
 {\cal L},_{k}Z^{\imath k}_{\alpha\beta\delta}(A) = 0 \;,
\end{eqnarray}
where
\begin{eqnarray*}
 D^{\gamma}_{\beta\delta\sigma}(A) &=&
(-1)^{{\varepsilon_\beta}{\varepsilon_\sigma}}
 \bigg({\cal F}^{\gamma}_{\beta\alpha}(A) {\cal F}^{\alpha}_{\delta\sigma}(A) +
{\cal F}^{\gamma}_{{\beta\delta},{\imath}}(A){\cal
R}^{\imath}_{\sigma}(A)\bigg)
\\
 & &+ {\rm cycl.perm.}(\beta\delta\sigma)\; ,
\end{eqnarray*}
\begin{eqnarray}
 Z^{\imath
k}_{\beta\delta\sigma}(A)&=&(-1)^{{\varepsilon_\beta}{\varepsilon_\sigma}}
 \bigg(M^{\imath k}_{\beta\alpha}(A){\cal F}^{\alpha}_{\delta\sigma}(A) +
 M^{\imath k}_{{\beta\delta},{\jmath}}(A){\cal R}^{\jmath}_{\sigma}(A)
\nonumber
\\
 & & -(-1)^{{\varepsilon_\beta}{\varepsilon_\imath}}{\cal
R}^{k}_{{\beta},{\jmath}}(A)
 M^{\imath\jmath}_{\delta\sigma}(A)
+(-1)^{{\varepsilon_k}({\varepsilon_\imath}+{\varepsilon_\beta})}{\cal
R}^{\imath}_{{\beta},{\jmath}}(A)
M^{k\jmath}_{\delta\sigma}(A)\bigg) \nonumber
\\
 & &\qquad+ {\rm cycl.perm.}(\beta\delta\sigma)\; ,
\end{eqnarray}
Given this
\begin{eqnarray}
 & &D^{\gamma}_{\beta\delta\sigma}(A) =
-(-1)^{{\varepsilon_{\beta\delta\sigma}}}D^{\gamma}_{\delta\beta\sigma}(A)=
-(-1)^{{\varepsilon_{\beta\delta\sigma}}}D^{\gamma}_{\beta\sigma\delta}(A)\;,
\nonumber
\\
 & &Z^{\imath k}_{\beta\delta\sigma}(A) =
-(-1)^{{\varepsilon_k}{\varepsilon_\imath}}Z^{k\imath}_{\beta\delta\sigma}(A)=
 -(-1)^{{\varepsilon_{\beta\delta\sigma}}}Z^{\imath k}_{\delta\beta\sigma}(A)=
\nonumber
\\
 & &\qquad -(-1)^{{\varepsilon_{\beta\delta\sigma}}}Z^{\imath
k}_{\beta\sigma\delta}(A)\;.
\end{eqnarray}
In Ref. 23 it is shown that Eq.(3.27) leads to the conclusion that there exist
the gauge algebra structural coeffitients
$Q^{\mu\imath}_{\beta\delta\sigma}(A)$
and ${\cal D}^{\imath\jmath k}_{\beta\delta\sigma}(A)$ such that
\begin{eqnarray*}
 & &\qquad Z^{\imath k}_{\beta\delta\sigma}(A) +
(-1)^{{\varepsilon_\mu}{\varepsilon_\imath}}
 {\cal R}^{k}_{\mu}(A)Q^{\mu\imath}_{\beta\delta\sigma}(A)
\\
 & & - (-1)^{({\varepsilon_\mu}+{\varepsilon_\imath}){\varepsilon_k}}{\cal
R}^{\imath}_{\mu}(A)Q^{\mu k}_{\beta\delta\sigma}(A) =
 -{\cal L},_{\jmath}(A){\cal D}^{\imath k \jmath}_{\beta\delta\sigma}(A)
\end{eqnarray*}
with
\begin{eqnarray}
 & &D^{\gamma}_{\beta\delta\sigma}(A) = {\cal L},_{k}(A)Q^{\gamma
k}_{\beta\delta\sigma}(A)\; .
\end{eqnarray}
Given this
\begin{eqnarray}
 & &Q^{\gamma k}_{\beta\delta\sigma}(A) =
-(-1)^{{\varepsilon_{\beta\delta\sigma}}}Q^{\gamma k}_{\delta\beta\sigma}(A)=
 -(-1)^{{\varepsilon_{\beta\delta\sigma}}}Q^{\gamma k}_{\beta\sigma\delta}(A)
\; ,\nonumber
\\
 & &{\cal D}^{\imath k \jmath}_{\beta\delta\sigma}(A)=
-(-1)^{{\varepsilon_k}{\varepsilon_\imath}}{\cal D}^{k
\imath\jmath}_{\beta\delta\sigma}(A)=
 -(-1)^{{\varepsilon_k}{\varepsilon_\jmath}}{\cal D}^{\imath\jmath
k}_{\beta\delta\sigma}(A) \; ,\nonumber
\\
 & &{\cal D}^{\imath k
\jmath}_{\beta\delta\sigma}(A)=-(-1)^{{\varepsilon_{\beta\delta\sigma}}}{\cal
D}^{\imath k \jmath}_{\delta\beta\sigma}(A)=
 -(-1)^{\varepsilon_{\beta\delta\sigma}}{\cal D}^{\imath k
\jmath}_{\beta\sigma\delta}(A)\;.
\end{eqnarray}
Then, by virtue of the gauge algebra relations (3.27), we find from definition
(2.27)
and the properties (3.15), (2.19), (2.29), (3.28) the following representation
for $F^a_3$
\begin{eqnarray}
 F^a_3 &=& \frac {1}{6}C^{*}_{\gamma b
c}D^{\gamma}_{\beta\delta\sigma}(A)T^{\sigma\delta\beta\{a b c\}}
\nonumber \\ &&                     
 +\frac {1}{6}\varepsilon^{c a}\bigg(C^{*}_{\gamma b c}+C^{*}_{\gamma c
b}\bigg)\bigg[
 {\cal F}^{\gamma}_{\delta\alpha}(A) {\cal F}^{\alpha}_{\beta\sigma}(A)
\nonumber \\ &&                     
 +2{\cal F}^{\gamma}_{{\delta\beta},{\imath}}(A){\cal
R}^{\imath}_{\sigma}(A)\bigg]
 (-1)^{\varepsilon_\beta \varepsilon_\delta}{\hat T}^{\sigma\delta}C^{\beta b}
\nonumber \\ &&                     
 -\frac {1}{12} A^{*}_{k b}A^{*}_{p c}(-1)^{\varepsilon_k}Z^{k
p}_{\beta\delta\sigma}(A)T^{\sigma\delta\beta\{a b c\}}
\nonumber \\ &&                     
 +\frac {1}{6}\varepsilon^{a b}A^{*}_{k b}A^{*}_{p c}(-1)^{\varepsilon_k}\bigg(
 M^{k p}_{\beta\alpha}(A){\cal F}^{\alpha}_{\delta\sigma}(A)+2M^{k
p}_{{\beta\delta},{\imath}}(A){\cal R}^\imath_\sigma (A)
\nonumber \\ &&                     
-2{\cal R}^p_{{\beta},{\imath}}(A)M^{k
\imath}_{\delta\sigma}(A)(-1)^{\varepsilon_k \varepsilon_\beta}\bigg)
 (-1)^{\varepsilon_\delta \varepsilon_\beta}{\hat T}^{\sigma\beta}C^{\delta c}
\nonumber \\ &&                     
 -\frac {1}{6}\varepsilon^{b c}A^{*}_{k b}A^{*}_{p c}(-1)^{\varepsilon_k}{\cal
R}^p_{{\beta},{\imath}}(A)M^{k \imath}_{\delta\sigma}(A)(-1)^{\varepsilon_\beta
(\varepsilon_\delta + \varepsilon_k)}
 {\hat T}^{\sigma\beta}C^{\delta a}
\nonumber \\ &&                     
 -\frac {1}{2} \bar{A}_k\bigg(N^k_{\beta\alpha}(A){\cal
F}^{\alpha}_{\delta\sigma}(A)+
 N^k_{\beta\sigma,\imath}(A){\cal R}^\imath_\delta (A)\bigg){\hat
T}^{\sigma\beta}C^{\delta a}(-1)^{\varepsilon_\delta (\varepsilon_\beta +
\varepsilon_\sigma)}
\nonumber \\ &&                     
 +\frac {1}{4}\bigg(2{\bar C}_{\gamma b}-B^{*}_{\gamma b}\bigg)\bigg[\bigg(
 {\cal F}^{\gamma}_{\sigma\alpha}(A) {\cal F}^{\alpha}_{\delta\beta}(A)
\nonumber \\ &&                     
 +2{\cal F}^{\gamma}_{{\sigma\delta},{\imath}}(A){\cal
R}^{\imath}_{\beta}(A)\bigg)
 (-1)^{\varepsilon_\sigma \varepsilon_\delta} - {\cal
F}^{\gamma}_{\delta\alpha}(A) {\cal F}^{\alpha}_{\sigma\beta}(A)\bigg]
 T^{\beta\sigma\{a b\}}B^\delta
\nonumber \\ &&                     
 +\frac {1}{4}\varepsilon^{a b}\bigg(2{\bar C}_{\gamma b}-B^{*}_{\gamma
b}\bigg)
 \bigg[{\cal F}^{\gamma}_{\sigma\alpha}(A) {\cal F}^{\alpha}_{\delta\beta}(A)
\nonumber \\ &&                     
 +2{\cal F}^{\gamma}_{{\sigma\delta},{\imath}}(A){\cal
R}^{\imath}_{\beta}(A)\bigg]
 (-1)^{\varepsilon_\sigma \varepsilon_\delta}{\hat T}^{\beta\sigma}B^\delta
\nonumber \\ &&                     
 +\frac {1}{4}A^{*}_{p b} {\bar A}_k \bigg[\bigg(M^{p k}_{\sigma\alpha}(A){\cal
F}^{\alpha}_{\delta\beta}(A)
 +2M^{p k}_{{\sigma\delta},{\imath}}(A){\cal R}^\imath_\beta (A)
\nonumber \\ &&                     
 +2{\cal R}^p_{\sigma,\imath}(A)M^{k
\imath}_{\delta\beta}(A)(-1)^{\varepsilon_k (\varepsilon_\sigma +
\varepsilon_p)}\bigg)
 (-1)^{\varepsilon_\sigma \varepsilon_\delta}
\nonumber \\ &&                     
 - M^{p k}_{\delta\alpha}(A){\cal
F}^{\alpha}_{\sigma\beta}(A)\bigg]T^{\beta\sigma\{a b\}}B^\delta
 +\frac {1}{4}\varepsilon^{a b}A^{*}_{p b} {\bar A}_k\bigg(M^{p
k}_{\sigma\alpha}(A){\cal F}^{\alpha}_{\delta\beta}(A)
\nonumber \\ &&                     
 +2{\cal R}^p_{\sigma,\imath}(A)M^{k
\imath}_{\delta\beta}(A)(-1)^{\varepsilon_k (\varepsilon_\sigma +
\varepsilon_p)}
\nonumber \\ &&                     
 +2M^{p k}_{{\sigma\delta},{\imath}}(A){\cal R}^\imath_\beta (A)\bigg)
 (-1)^{\varepsilon_\sigma \varepsilon_\delta}{\hat T}^{\beta\sigma}B^\delta
\nonumber \\ &&                     
 -\frac {1}{2}{\bar A}_p{\bar A}_k{\cal R}^k_{{\sigma},{\imath}}(A)M^{p
\imath}_{\delta\beta}(A)
 C^{\beta a}B^\delta B^\sigma(-1)^{\varepsilon_\sigma (\varepsilon_p
+1)+\varepsilon_\delta} \; ,
\end{eqnarray}
Solution to Eqs. (3.11) with the right-hand side (3.32) is sought in a way
similar to the one given in the case of $S_2$, that is, by consideration of
each summond present in $F^a_3$. To this end, it is necessary to employ all
the gauge algebra structural relations (3.1), (3.14), (3.27), (3.30), as well
as the properties of structural coeffitients.

The functional $N^\imath_{\alpha\beta}(A)$, not being a structural coeffitient,
along with the functional $N^\imath_{\alpha\beta,_\jmath}(A)$ are treated
with the help of the differential consequence of the property (3.1)
\begin{eqnarray}
 {\cal L},_{\jmath k}(A){\cal R}^k_\delta (A)(-1)^{\varepsilon_\jmath
\varepsilon_\delta}
 = - {\cal L},_{k}(A){\cal R}^k_{\delta ,\jmath} (A) \; .
\end{eqnarray}

Omitting details of cumbersome algebraic calculations, we give here the
resultant form for $S_3$
\begin{eqnarray}
 S_3 &=& -\frac {1}{6}C^{*}_{\gamma a b}A^{*}_{\jmath
c}Q^{\gamma\jmath}_{\beta\delta\sigma}(A)
 T^{\sigma\delta\beta\{a b c\}}(-1)^{\varepsilon_\beta \varepsilon_\sigma}
\nonumber \\ &&                     
 -\frac {1}{6} C^{*}_{\gamma a b}\bar
{A}_{\jmath}Q^{\gamma\jmath}_{\beta\delta\sigma}(A)
 T^{\beta\sigma\{a b\}}B^\delta (-1)^{\varepsilon_\beta \varepsilon_\delta}
\nonumber \\ &&                     
 +\frac{1}{36}A^{*}_{\jmath b}A^{*}_{\imath c}A^{*}_{k a}{\cal D}^{\jmath\imath
k}_{\beta\delta\sigma}(A)
 T^{\sigma\delta\beta\{a b c\}}(-1)^{\varepsilon_\beta (\varepsilon_\sigma
+\varepsilon_\delta) + \varepsilon_\imath}
\nonumber \\ &&                     
 -\frac {1}{6}\bigg(2{\bar C}_{\gamma b}-B^{*}_{\gamma b}\bigg)A^{*}_{\jmath c}
 Q^{\gamma\jmath}_{\beta\delta\sigma}(A)T^{\beta\sigma\{c b\}}B^\delta
(-1)^{\varepsilon_\beta \varepsilon_\delta  +\varepsilon_\gamma}
\nonumber \\ &&                     
 -\frac {1}{6}\bigg(2{\bar C}_{\gamma b}-B^{*}_{\gamma b}\bigg)\bigg[
 {\cal F}^{\gamma}_{\delta\alpha}(A) {\cal F}^{\alpha}_{\beta\sigma}(A)
\nonumber \\ &&                     
 +2{\cal F}^{\gamma}_{{\delta\beta},{\imath}}(A){\cal
R}^{\imath}_{\sigma}(A)\bigg]
 (-1)^{\varepsilon_\beta \varepsilon_\delta}{\hat T}^{\sigma\delta}C^{\beta b}
\nonumber \\ &&                     
 +\frac {1}{12}A^{*}_{\jmath b}A^{*}_{p c}\bar{A}_{k}{\cal D}^{\jmath p
k}_{\beta\delta\sigma}(A)
 T^{\beta\sigma\{b c\}}B^\delta(-1)^{\varepsilon_\beta \varepsilon_\delta +
\varepsilon_\jmath}
\nonumber \\ &&                     
 +{\bar A}_p{\bar A}_k{\cal R}^k_{{\sigma},{\imath}}(A)M^{p
\imath}_{\delta\beta}(A)
(-1)^{\varepsilon_\sigma(\varepsilon_p +\varepsilon_\delta)}{\hat
T}^{\beta\sigma}B^{\delta}
\nonumber \\ &&                     
 +\frac {1}{6}A^{*}_{p b}{\bar A}_k\bigg(2{\cal R}^p_{{\beta},{\imath}}(A)M^{k
\imath}_{\delta\sigma}(A)(-1)^{\varepsilon_k (\varepsilon_\beta +
\varepsilon_p)}
\nonumber \\ &&                     
 +4{\cal R}^k_{{\beta},{\imath}}(A)M^{p
\imath}_{\delta\sigma}(A)(-1)^{\varepsilon_p \varepsilon_\beta}
 -M^{p k}_{\beta\alpha}(A){\cal F}^{\alpha}_{\delta\sigma}(A)
\nonumber \\ &&                     
 -2M^{p k}_{{\beta\delta},{\imath}}(A){\cal R}^\imath_\sigma (A)\bigg)
 {\hat T}^{\sigma\beta}C^{\delta b}(-1)^{\varepsilon_\beta
\varepsilon_\delta}\; .
\end{eqnarray}
Thus, in both the standard and Sp(2)-covariant formulations of Lagrangian
BRST quantization, the generating equations give rise to the gauge algebra
structural relations for every order in power series of ghost and
auxiliary fields $C^{\alpha a}$, $B^\alpha$. The boson functional
\begin{eqnarray}
 S^{(0)} &=& {\cal L}(A) +S_1(\Phi^A,\Phi^{*}_{A a},{\bar \Phi}_A)
 +S_2 (\Phi^A,\Phi^{*}_{A a},{\bar \Phi}_A)
\nonumber \\ &&                     
  +S_3 (\Phi^A,\Phi^{*}_{A a},{\bar \Phi}_A)+O(C^{4-m} B^m),
\nonumber \\ &&                     
 \qquad m=0,1,2,3,4.
\end{eqnarray}
satisfies the generating equations (3.8) up to the third order inclusively.
Since arbitrariness in solution to Eqs. (3.8) is completely described in
Refs. 15, 17, we omit the discussion of this question.

Consider a particular case of gauge theories of rank 1 with a closed algebra,
i. e. let
\begin{eqnarray*}
 M^{p \imath}_{\delta\beta}(A)=Q^{\gamma\jmath}_{\beta\delta\sigma}(A)=
 D^{\gamma}_{\beta\delta\sigma}(A)={\cal D}^{\jmath\imath
k}_{\beta\delta\sigma}(A)=0
\end{eqnarray*}
Then, the functional (3.34) takes on the form
\begin{eqnarray}
 S^{(0)}(\Phi^A,\Phi^{*}_{A a},{\bar \Phi}_A)={\cal L}(A)+\Phi^{*}_{A a}X^{A a}
 + {\bar \Phi}_A Y^A \;\; ,
\end{eqnarray}
where
\begin{eqnarray*}
 X^{A a} &=& \bigg(X^{\imath a}_1 , X^{\alpha a}_2 , X^{\alpha a b}_3\bigg)\;
,\;
 Y^{A} = \bigg(Y^{\imath}_1 , Y^{\alpha}_2 , Y^{\alpha a}_3\bigg)\;\; ,
\end{eqnarray*}
given this
\begin{eqnarray}
 X^{\imath a}_1&=& {\cal R}^\imath_\alpha (A)C^{\alpha a}\;\;,
\nonumber \\
 X^{\alpha a}_2&=& -\frac {1}{2}{\cal F}^\alpha_{\gamma\beta}(A)B^\beta
C^{\gamma a}
 -\frac{1}{12} (-1)^{\varepsilon_\beta}\bigg({\cal
F}^{\alpha}_{\gamma\sigma}(A) {\cal F}^{\sigma}_{\beta\rho}(A)
\nonumber \\ &&
 +2{\cal F}^{\alpha}_{{\gamma\beta},{\imath}}(A){\cal
R}^{\imath}_{\rho}(A)\bigg)
 C^{\rho b}C^{\beta a}C^{\gamma c} \varepsilon_{c b}\; ,
\nonumber \\
 X^{\alpha a b}_3&=& -\varepsilon^{a b}B^\alpha
-\frac{1}{2}(-1)^{\varepsilon_\beta}
 {\cal F}^\alpha_{\beta\gamma}(A)C^{\gamma b}C^{\beta a}\; ,
\nonumber \\
 Y^\imath_1&=& {\cal R}^\imath_\alpha
(A)B^\alpha+\frac{1}{2}(-1)^{\varepsilon_\alpha}
 {\cal R}^\imath_{\alpha ,\jmath} (A){\cal R}^\jmath_\beta (A)C^{\beta
b}C^{\alpha a}
 \varepsilon_{a b}\;,
\nonumber \\
 Y^\alpha_2 &=& 0 \;,\;Y^{\alpha a}_3 = -2 X^{\alpha a}_2\; .
\end{eqnarray}
The result (3.36), (3.37) coincides with the exact solution to the generating
equations for $S^{(0)}$ obtained in Ref. 15 in the case of gauge theories of
rank 1 with a closed algebra.\\[1mm]
\setcounter{sct}{4}
\setcounter{equation}{0}
\\{\bf 4. Sp(2)-covariant Quantization of the Yang-Mills Theory}\\[0.1mm]

 It may not be out of place to illustrate general results and relations of this
 paper on a basis of a simple example of a concrete gauge theory. To do this
 we consider the Yang-Mills theory described by the action
\begin{eqnarray}
 {\cal L}={\cal L}(A^{\mu m})=-\frac{1}{4}\int d^4x\;G_{\mu\nu}^mG^
 {\mu\nu m}.
\end{eqnarray}
 The Yang-Mills field $A^{\mu m}(x)$ is defined on the {\mbox Minkovsky} space
 and assumes its values in the adjoint representation of a semisimple compact
 group. The field strength $G_{\mu\nu}^m$ has the form
\[
 G_{\mu\nu}^m=\partial_\mu A_\nu^m-\partial_\nu A_\mu^m+f^{mnl}A_\mu^n A_\nu^l,
\]
 where the Greek subscripts $\mu$, $\nu$ refer to the {\mbox Minkovsky} space,
 while the Roman subscripts $k$, $l$, $m$, $n$, $p$, refer to the internal
 symmetry group indices. The Yang-Mills fields $A^{\mu m}$ play the role of
 initial fields $A^\imath$. The action (4.1) is invariant under the
 gauge transformations \[ \delta A_\mu^m(x)={\cal
 D}_\mu^{mn}(x)\xi^n(x)=\int d^4y\;{\cal R}_\mu^ {mn}(x;y)\xi^n(y), \]
 where
\begin{eqnarray}
 {\cal D}_\mu^{mn}=\delta^{mn}\partial_\mu+f^{mln}A_\mu^l
\end{eqnarray}
 is the covariant derivative and ${\cal R}_\mu^{mn}(x;y)$ are the generators
 of the gauge transformations
\[
 {\cal R}_\mu^{mn}(x;y)={\cal D}_\mu^{mn}\delta(x-y).
\]
 In (4.2) $f^{lmn}$ are the structural constants associated with the symmetry
 group (the interaction constant is absorbed into $f^{lmn}$), $\xi^n(y)$ are
 arbitrary functions. The condensed indices $\imath$, $\alpha$ for the theory
 in question are understood in the form
\[
 \imath=(\mu,m,x),\;\;\;\;\alpha=(m,x).
\]
 The structural coeffitients ${\cal F}^\gamma_{\alpha\beta}$ arising in the
relations
 (3.14) are defined by the group structural constants $f^{lmn}$ in the form
 $(\alpha=(m,x),\beta=(n,y),\gamma=(l,z))$
\[
 {\cal F}^\gamma_{\alpha\beta}=f^{lmn}\delta(x-z)\delta(y-z),
\]
 while $M^{\imath\jmath}_{\alpha\beta}=0$. The higher gauge algebra structural
 coeffitients $Q^{\imath\alpha}_{\beta\gamma\delta}$, ${\cal D}^{\imath\jmath
 k}_{\beta\gamma\delta}$ also assume zero value. The total configuration space
of the theory
 is defined by the set of fields
\begin{eqnarray}
 \Phi^A=(A^{\mu m},B^m, C^{ma})
\end{eqnarray}
 and therefore, the set of antifields $\Phi^{\ast}_{Aa}$ and $\bar{\Phi}_A$
have
 the form
\[
 \Phi^{\ast}_{Aa}=(A^{\ast m}_{\mu a},B^{\ast m}_a,C^{\ast m}_{ab}),\;\;\;\;
 \bar{\Phi}_A=(\bar{A}_\mu^m,\bar{B}^m,\bar{C}^m_a).
\]
 Given this, the Grassmann parity and the new ghost number assume the values
 $(\varepsilon_i=\varepsilon_\alpha=0)$
\[
 \varepsilon(A^{\mu m})=\varepsilon(B^m)=0,\;\;\;\;\varepsilon(C^{ma})=1,
\]
\[
 \varepsilon(A^{\ast m}_{\mu_a})=\varepsilon(B^{\ast m}_a)=1,\;\;\;\;
 \varepsilon(C^{\ast m}_{ab})=0,
\]
\[
 \varepsilon(\bar{A}_\mu^m)=\varepsilon(\bar{B}^m)=0,\;\;\;\;
 \varepsilon(\bar{C}^m_a)=1,
\]
\[
 {\rm ngh}(A^{\mu m})=0,\;\;\;\;{\rm ngh}(C^{am})=1,\;\;\;\;{\rm ngh}(B^m)=2,
\]
\[
 {\rm ngh}(A^{\ast m}_{\mu a})=-1,\;\;\;\;{\rm ngh}(C^{\ast m}_{ab})=-2,
 \;\;\;\;{\rm ngh}(B^{\ast m}_a)=-3,
\]
\[
 {\rm ngh}(\bar{A}_\mu^m)=-2,\;\;\;\;{\rm ngh}(\bar{C}^m_a)=-3,\;\;\;\;
 {\rm ngh}(\bar{B}^m)=-4.
\]
 By virtue of the manifest structure (4.1)--(4.3) of the theory in question,
 the solutions to the generating equations for $S_1$, $S_2$, $S_3$ are
 representable in the following local form
\[
 S_1=\int d^4x\bigg\{A^{\ast m}_{\mu a}{\cal D}^{\mu mn}C^{na}+\bar{A}_\mu^m
 {\cal D}^{\mu mn}B^n-C^{\ast m}_{ab}B^m\varepsilon^{ab}\bigg\},
\]
\begin{eqnarray}
 S_2&=&\int d^4x\bigg\{\bigg(\bar{C}^m_a-\frac{1}{2}B^{\ast m}_a\bigg)f^{mnl}
 B^lC^{na}\nonumber\\&&
 -\frac{1}{2}C^{\ast m}_{ab}f^{mnl}C^{lb}C^{na}-\frac{1}{2}\bar{A}_\mu^m
 f^{mnl}C^{na}{\cal D}^{\mu lk}C^{kb}\varepsilon_{ab}\bigg\},
\end{eqnarray}
\[
 S_3=-\frac{1}{12}\int d^4x\bigg\{\bigg(B^{\ast m}_a-2\bar{C}^m_a\bigg)f^{mnl}
 f^{lkp}C^{pb}C^{ka}C^{nc}\varepsilon_{cb}\bigg\}.
\]
 Since the Yang-Mills theory belongs to the theories of rank 1, the functional
 $S={\cal L}+S_1+S_2+S_3$ is an exact solution to the generating equations.
 Moreover, the antisymmetry properties of the structural constants $f^{lmn}$
 lead to the fact that the functional $S$ is also a solution to Eqs.~(3.3).

 Consider for the theory in question the generating functional of Green's
 functions $Z(J)$ (3.6). Choose the gauge functional $F$ in the form
\begin{eqnarray}
 F=-\frac{\alpha}{2}\int d^4x\;A_\mu^mA^{\mu m}.
\end{eqnarray}
 Then, integrating in the functional integral (3.6) over the variables
 $\lambda^A$, $\Pi^{Aa}$, $\bar{\Phi}_A$, $\Phi^{\ast}_{Aa}$ and
 taking (4.4), (4.5) into account, we obtain the following
 representation for the generating functional $Z(J)$ \begin{eqnarray}
 Z(J)=\int d\Phi\;\exp\bigg\{\frac{i}{\hbar}\bigg({\cal L}+S_{\rm GH}(\Phi)+
 S_{\rm GF}(\Phi)+J_A\Phi^A\bigg)\bigg\},
\end{eqnarray}
 where the following notations are used
\[
 S_{\rm GH}=\frac{\alpha}{2}\int d^4x\;\partial^\mu C^{ma}{\cal D}_\mu^{mn}
 C^{nb}\varepsilon_{ab},
\]
\[
 S_{\rm GF}=\alpha\int d^4x\;B^m\partial^\mu A^m_\mu.
\]
\hspace*{\parindent} Note that the integrand in Eq. (4.6) for $J_A=0$ is
 invariant under the extended BRST symmetry transformations of the form
\[
 \delta A_\mu^m={\cal D}_\mu^{mn}C^{na}\mu_a,
\]
\[
 \delta B^m=-\frac{1}{2}\bigg(f^{mnl}B^lC^{na}+\frac{1}{6}f^{mnl}f^{lkp}C^{pb}
 C^{ka}C^{nc}\varepsilon_{cb}\bigg)\mu_a,
\]
\[
 \delta C^{ma}=\bigg(\varepsilon^{ab}B^m-\frac{1}{2}f^{mnl}C^{la}C^{nb}\bigg)
 \mu_b.
\]
\hspace*{\parindent} Now consider the Yang-Mills theory in the
 {\mbox Hamiltonian} version of Sp(2)-covariant quantization. Note to this end
 that the corresponding dynamical system is described in the initial
 phase space $\eta$ ($x^\mu=(x^0, \vec{x})$, the spatial indices are denoted
 $i$, $j$:  $\mu=(0,i)$)
 \[
 \eta=(p_\imath,q^\imath)=(\Pi_i^m,A^{im}),\;\;\;\;\imath=(i,m,\vec{x}),
 \;\;\;\;\varepsilon(A^{im})=0
\]
 by the classical Hamiltonian $H_0$
\[
 H_0=\int d^3x\bigg\{-\frac{1}{2}\Pi_i^m\Pi^{im}+\frac{1}{4}F_{ij}^mF^{ijm}
 \bigg\}
\]
 and by the set of lineary independent constraints $T_\alpha$ ($\alpha=(m,\vec
 {x})$)
\[
 T_\alpha=T^m\equiv{\cal D}_i^{mn}\Pi^{in},
\]
 with the following involution relaitons
\[
 \{T^m(x),T^n(y)\}=\int d^3z\;f^{lmn}T^l(z)\delta(\vec{x}-\vec{z})
 \delta(\vec{y}-\vec{z}),\;\;\;\;x^0=y^0=z^0,
\]
\[ \{T^m(x),H_0\}=0. \]
 Hence the structural coeffitients $U^\gamma_{\alpha\beta}$, arising in Eqs.
 (2.1) have the form ($\alpha=(m,\vec{x})$, $\beta=(n,\vec{y})$, $\gamma=(l,
 \vec{z})$)
\begin{eqnarray}
 U^\gamma_{\alpha\beta}=f^{lmn}\delta(\vec{x}-\vec{z})\delta(\vec{y}-\vec{z}),
\end{eqnarray}
 whereas $V^\beta_\alpha=0$. Given this, the higher gauge algebra structural
 coeffitients $E^{\gamma\delta}_{\alpha\beta}$, $E^{\delta\rho}_
 {\alpha\beta\gamma}$, $E^{\delta\rho\sigma}_{\alpha\beta\gamma}$ are equal
 to zero.

 The extended phase space $\Gamma$ for the dynamical system in question has
 the form
\begin{eqnarray}
 \Gamma=(P_A,Q^A)=(\eta;{\cal P}^m_a,C^{ma};\lambda^m,\pi^m),
\end{eqnarray}
 where the Grassmann parity and the new ghost number of the variables belong
 to $\Gamma$ are as follows
\[
 \varepsilon(C^{ma})=1,\;\;\;\;\varepsilon(\pi^m)=0,
\]
\[
 {\rm ngh}(C^{ma})=1,\;\;\;\;{\rm ngh}(\pi^m)=2.
\]
 The explicit form of the gauge algebra structural coeffitients (4.7) and the
 manifest structure of the extended phase space $\Gamma$ (4.8) enable us, with
 allowance made for Eqs. (2.44), to give solutions to the generating
 equations for ${\cal H}$, $\Omega^a$ \[ {\cal H}=H_0, \]
\begin{flushright}(4.9)\end{flushright}
\setcounter{equation}{9}
\begin{eqnarray*}
 \Omega^a&=&\int d^3x\bigg\{C^{ma}{\cal D}_i^{mn}\Pi^{in}
 +\varepsilon^{ab}{\cal P}^m_b\pi^m+\frac{1}{2}{\cal P}^l_bf^{lmn}C^{na}
 C^{mb}\\&&
 -\frac{1}{2}\lambda^lf^{lmn}C^{na}\pi^m-\frac{1}{12}\lambda^
 lf^{lmn}f^{nkp}C^{pa}C^{kb}C^{mc}\varepsilon_{bc}\bigg\}.
\end{eqnarray*}
\hspace*{\parindent} Now consider, by virtue of (4.9), the generating
 functional of Green's functions $Z(I)$ (2.8) in the Hamiltonian version of
 Sp(2)-covariant quantization. To this end we choose the boson gauge function
$\Phi$ in Eq.~
 (2.6), determining the unitarizing {\mbox Hamiltonian} $H$, in the form
\[
 \Phi=\int d^3x\bigg\{\frac{\alpha}{2}A_i^mA^{im}-\frac{1}{2\alpha}\lambda^m
 \lambda^m\bigg\}.
\]
 Then, integrating in the functional integral (2.8) over the momenta
 ${\cal P}^m_a$ and assuming the corresponding sources to be equal to zero,
 we obtain, with allowance made for the notations of the form
\[
 A_0^m\equiv-\alpha^{-1}\lambda^m,\;\;\;\;B^m\equiv\pi^m,
\]
 the following representation for the generating functional of Green's
 functions (2.8)
\[
 Z(J)=\int d\Phi\;\exp\bigg\{\frac{i}{\hbar}\bigg({\cal L}+S_{\rm GH}+
 S_{\rm GF}+J_A\Phi^A\bigg)\bigg\}.
\]
 Here $\Phi^A$ and $J_A$ coincide with the sets of fields of the total
 configuration space (4.3) and the corresponding sources respectively.
 The functional ${\cal L}$ is the classical action (4.1), whereas the
 functionals $S_{\rm GH}$, $S_{\rm GF}$ are defined in Eq. (4.6). Hence we
 conclude that for the fields of the total configuration space in the theory
 (4.1)--(4.2), the generating functionals of Green's functions of both the
 {\mbox Lagrangian} and {\mbox Hamiltonian} versions of Sp(2)-covariant
 quantization coincide.\\[0.5mm]
\\{\bf Acknowledgment}\\[0.1mm]

 The authors are grateful to I. A. Batalin and I. V. Tyutin for useful
 discussions. The work is supported in part by the International Science
 Foundation, grant RI~1000 and by the Russian Foundation for Fundamental
 Research, project No. 94--02--03234.
\newpage
{\bf References}
\begin{itemize}
\item[{1.}] B. de Wit and J. W. van Holten, \underline{Phys. Lett.} B79
           (1978) 389.
\item[{2.}] I. A. Batalin and G. A. Vilkovisky, \underline{Phys. Lett.}
            B102 (1981) 27.
\item[{3.}] I. A. Batalin and G. A. Vilkovisky, \underline{Phys. Rev.} D28
           (1983) 2576.
\item[{4.}] E. S. Fradkin and G. A. Vilkovisky, \underline{Phys. Lett.} B55
           (1975) 224.
\item[{5.}] I. A. Batalin and G. A. Vilkovisky, \underline{Phys. Lett.} B69
           (1977) 309.
\item[{6.}] E. S. Fradkin and T. E. Fradkina, \underline{Phys. Lett.} B72
           (1978) 343.
\item[{7.}] C. Becchi, A. Rouet and R. Stora, \underline{Commun. Math. Phys.}
            42 (1975) 127.
\item[{8.}] I. V. Tyutin, preprint Lebedev Inst. No. 39, 1975.
\item[{9.}] N. Nakanishi and I. Ojima, \underline{Z. Phys.} C6 (1980) 255.
\item[{10.}] L. Alvarez-Gaume and L. Baulieu, \underline{Nucl. Phys.} B212
          (1983) 255.
\item[{11.}] S. Hwang, \underline{Nucl. Phys.} B231 (1984) 386.
\item[{12.}] V. P. Spiridonov, \underline{Nucl. Phys.} B308 (1988) 527.
\item[{13.}] G. Curci and R. Ferrari, \underline{Phys. Lett.} B63 (1976) 91.
\item[{14.}] I. Ojima, \underline{Progr. Theor. Phys.} 63 (1979) 625.
\item[{15.}] I. A. Batalin, P. M. Lavrov and I. V. Tyutin, \underline{J. Math.
          Phys.} 31 (1990) 1487.
\item[{16.}] I. A. Batalin, P. M. Lavrov and I. V. Tyutin, \underline{J. Math.
          Phys.} 32 (1991) 532.
\item[{17.}] I. A. Batalin, P. M. Lavrov and I. V. Tyutin, \underline{J. Math.
          Phys.} 32 (1991) 2513.
\item[{18.}] I. A. Batalin, P. M. Lavrov and I. V. Tyutin, \underline{J. Math.
          Phys.} 31 (1990) 6.
\item[{19.}] I. A. Batalin, P. M. Lavrov and I. V. Tyutin, \underline{J. Math.
          Phys.} 31 (1991) 2708.
\item[{20.}] I. A. Batalin, P. M. Lavrov and I. V. Tyutin, \underline{Int. J.
Mod.
          Phys.} A6 (1991) 3599.
\item[{21.}] B. S. De Witt, \underline{Dynamical Theory of Groups and Fields}
          (Gordon and Breach, New York, 1965).
\item[{22.}] E. S. Fradkin, proceedings of the X Winter School of Theoretical
          Physics in Karpacs, No. 207, 1973.
\item[{23.}] I. A. Batalin and G. A. Vilkovisky, \underline{J. Math. Phys.}
          26 (1985) 172.
\end{itemize}
\end{document}